\documentclass[11pt]{article}

\usepackage[colorlinks=true,allcolors=black]{hyperref}

\usepackage{scicite}

\usepackage{amsmath}
\usepackage{amsfonts}
\usepackage{amssymb}
\usepackage{amsthm}
\usepackage{graphicx}
\usepackage{enumerate}
\usepackage[table,dvipsnames]{xcolor}

\usepackage{sectsty}
\sectionfont{\color{Mahogany}}

\newcommand{\id}{\mathrm d}
\usepackage{bm}\newcommand{\vc}{\bm}
\newcommand{\bnabla}{\pmb\nabla}

\newcommand{\re}{\mbox{Re}}
\newcommand{\im}{\mbox{Im}}

\newcommand{\Dmax}{D_{m}}
\theoremstyle{definition}

\newcommand{\edit}[1]{{\color{black}#1}}

\usepackage{xr}

\newenvironment{sciabstract}{
\begin{quote} \bf}
{\end{quote}}

\title{A variational approach to probing extreme events in turbulent dynamical systems}

\author
{Mohammad Farazmand,${}^{\ast}$ Themistoklis P. Sapsis${}^{\ast}$\\
\\
\normalsize{Department of Mechanical Engineering, Massachusetts Institute of Technology}\\
\normalsize{77 Massachusetts Avenue, Cambridge MA 02139-4307}\\
\\
\normalsize{$^\ast$To whom correspondence should be addressed; E-mail: mfaraz@mit.edu, sapsis@mit.edu.}
}

\date{}

\begin{document} 




\maketitle 



\begin{sciabstract}
Extreme events are ubiquitous in a wide range of dynamical systems, including turbulent fluid flows, nonlinear waves, large scale networks and biological systems. Here, we propose a variational framework for probing conditions that trigger intermittent extreme events in high-dimensional nonlinear dynamical systems. 
We seek the triggers as the probabilistically feasible solutions of an appropriately constrained optimization problem,
where the function to be maximized is a system observable exhibiting intermittent extreme bursts.
The constraints are imposed to ensure the physical admissibility of the optimal solutions, i.e., 
significant probability for their occurrence under the natural flow of the dynamical system. 
We apply the method to a body-forced incompressible Navier--Stokes
equation, known as the Kolmogorov flow. We find that the intermittent bursts of the energy dissipation 
are independent of the external forcing and are instead caused by the spontaneous transfer of energy from large scales to the mean flow via nonlinear triad interactions. The global maximizer of the corresponding variational problem identifies the responsible triad, hence providing a precursor for the occurrence of extreme dissipation events. Specifically, monitoring the energy transfers within this triad, allows us to develop a data-driven short-term predictor for the intermittent bursts of energy dissipation. We assess the performance of this predictor through direct numerical simulations.
\end{sciabstract}


\section*{Introduction}

A plethora of dynamical systems exhibit intermittent behavior manifested
through sporadic bursts in the time series of their observables. These extreme 
events produce values of the observable that are several standard deviations 
away from its mean, resulting in heavy tails of the corresponding probability distribution. 
Important examples include climate phenomena~\cite{Neelin1998, Thual2016}, 
rogue waves in oceanic and optical systems~\cite{Dysthe08, onorato13,dematteis2017} and
large deviations in turbulent flows~\cite{donzis2010,yeung2015,qi15}.
Since such extreme phenomena typically have dramatic consequences, their quantification and prediction is of great interest.

Significant progress has been made in the computation of extreme statistics, both through
direct numerical simulations and through indirect methods such as the theory of large
deviations (see~\cite{touchette2009} for a review).
While these methods estimate the probability distribution of the extreme events, 
they do not inform us about the underlying mechanisms that lead to the extremes nor are they 
capable of predicting individual extreme events.

For systems operating near an equilibrium or systems that are 
nearly integrable, perturbative methods 
have been successful in identifying the resonant interactions that cause the
extreme events~\cite{jones95,haller99}. For systems which are not perturbations from such trivial limits, 
a general framework for probing the transition mechanism to
extreme states is missing. These systems, such as turbulent fluid flows and
water waves, are also typically high-dimensional and nonlinear where the nonlinearities create 
a complex network of interdependent interactions among many 
degrees of freedom~\cite{sapsis11a,Majda2015,cousins_sapsis,cousinsSapsis2015_JFM,brunton2016,Farazmand2017}.

Here, we propose a variational framework to probe the underlying conditions that
lead to extreme events in such high-dimensional complex systems. 
More specifically, we derive precursors of extreme events
as the solutions of a finite-time constrained optimization problem. The functional
to be maximized is the observable whose time series exhibit the extreme events.
The constraints are designed to ensure that the 
triggers belong to the system attractor and therefore 
reflect physically relevant phenomena. If the life-time of the extreme events are short compared to the typical  dynamical time 
scales of the system, the finite-time optimization
problem can be replaced with its instantaneous counterpart. 
For the instantaneous problem, we derive the Euler-Lagrange 
equations that can be solved numerically using Newton-type iterations.

We apply the variational framework to the Kolmogorov flow, 
a two-dimensional Navier--Stokes equation driven by a monochromatic body forcing. 
At sufficiently high Reynolds numbers, this flow is known to exhibit intermittent bursts of 
energy dissipation rate~\cite{faraz_adjoint,PRE2016}.  We first show that these extreme bursts are 
due to the internal transfers of energy through nonlinearities, as opposed to phase locking
with the external forcing. Because of the high number of involved degrees of freedom and their complex interactions, 
deciphering the modes responsible for the extreme energy dissipation rate is not straightforward.
The optimal solution to our variational method, however, isolates the triad interaction
responsible for this transfer of energy. Monitoring this triad along trajectories of the Kolmogorov flow,
we find that, on the onset of the extreme bursts,
the energy is transferred spontaneously from a large-scale Fourier mode 
to the mean flow, leading to growth of the
energy input rate and consequently the energy dissipation rate.

We then utilize the derived large-scale mode as a predictor for the extreme events. Specifically, by tracking the energy of this mode we develop a data-driven short-term predictor of intermittency in the Kolmogorov flow. 
We assess the effectiveness of the prediction scheme on extensive direct numerical simulations 
by explicitly quantifying its success rate, as well as the false positive and false negative rates. 

\section*{Results}
\subsection*{Variational formulation of extreme events}
Consider the general evolution equations,
\begin{subequations}
\begin{equation}
\partial_t\vc u = \mathcal N(\vc u),
\label{eq:pde_1}
\end{equation}
\begin{equation}
\mathcal K(\vc u)=0,
\label{eq:pde_2}
\end{equation}
\begin{equation}
\vc u(\cdot, t_0)=\vc u_0(\cdot),
\label{eq:pde_3}
\end{equation}
\label{eq:pde}%
\end{subequations}
where $\vc u:\Omega\times \mathbb R^+\to\mathbb R^d$ belongs to an 
appropriate function space $X$ and completely 
determines the state of the system. The initial condition
$X\ni\vc u_0:\Omega\to \mathbb R^d$ is specified at the time $t_0$ and
$\Omega\subset \mathbb R^d$ is an open bounded domain. 
The differential operators $\mathcal N$ and $\mathcal K$ 
can be potentially nonlinear.
A wide range of physical models can be written as a set of partial differential 
equations (PDEs) as in~\eqref{eq:pde}. 
For instance, for incompressible 
fluid flows, ~\eqref{eq:pde_1} is the momentum equation 
and~\eqref{eq:pde_2} is the incompressibility condition where
$\mathcal K(\vc u)=\bnabla\cdot\vc u$.
For simplicity, we will denote a trajectory of~\eqref{eq:pde} by $\vc u(t)$.

\begin{figure}[h]
\centering
\includegraphics[width=.7\textwidth]{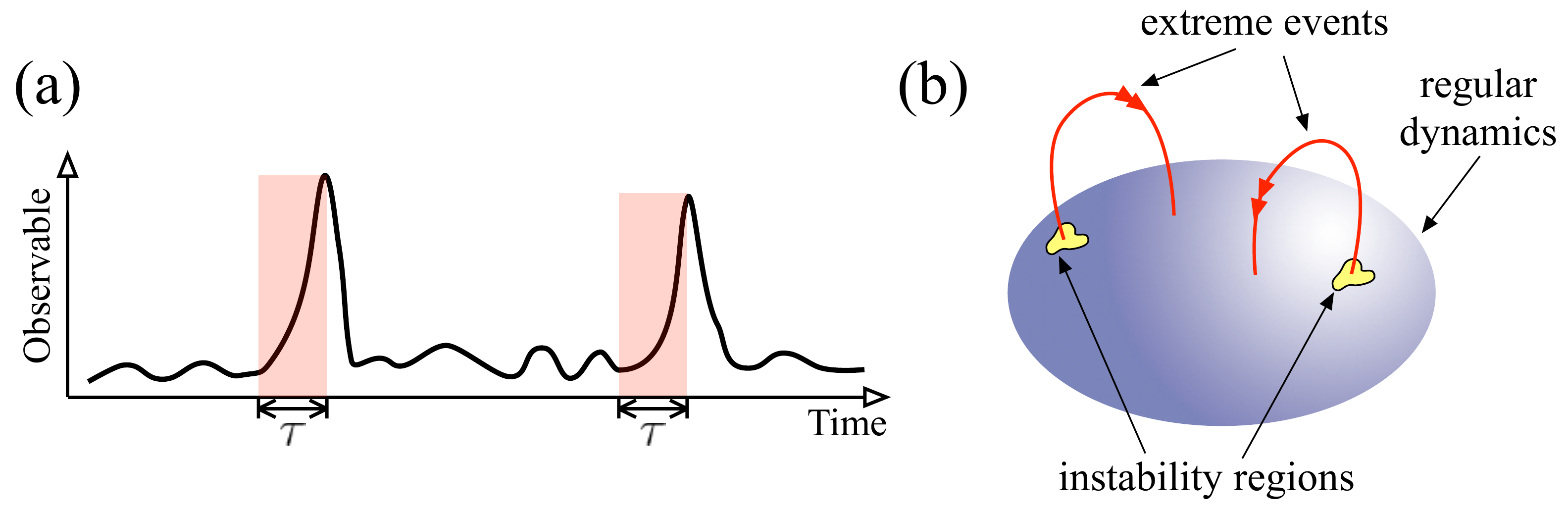}
\caption{
(a) A depiction of intermittent bursts of an observable.
The highlighted regions mark an approximation of the growth
phase of the extreme events.
(b) In the the state space, extreme events are viewed as fast 
excursions away from the background attractor (blue ball) due to small regions of stability.}
\label{fig:schem_FTgrowth}
\end{figure}

Let $I:X\rightarrow \mathbb R$ denote an observable whose 
time series $I(\vc u(t))$ along a typical trajectory $\vc u(t)$ 
exhibits intermittent bursts (see figure~\ref{fig:schem_FTgrowth}a).
Drawing upon the near-integrable case, we view the system as 
consisting of a background chaotic attractor which has small regions of 
instability (see figure~\ref{fig:schem_FTgrowth}b).
Once a trajectory reaches an instability region, it is momentarily repelled away from the 
background attractor, resulting in a burst in the time series of the observable. 
Our goal here is to probe the instability region(s) by utilizing a combination of observed data 
from the system and the governing equations of the system.
We also require the instability regions to have non-zero probability of occurrence 
under the natural flow of the dynamical system.
This constraint is of particular importance since it excludes `exotic' states with extreme growth of $I$
but with negligible probability of being observed in practice (see the constraint $\vc C(\vc u_0)=\vc c_0$ in 
equation~\eqref{eq:FTopt_const} below).

We formulate this task as a constrained optimization problem. 
Assume that there is a typical timescale $\tau\in \mathbb R^+$ over which the 
bursts in the observable $I$ develop (see figure~\ref{fig:schem_FTgrowth}a). 
We seek initial conditions $\vc u_0$ whose associated observable 
$I(\vc u(t))$ attains a maximal growth within time $\tau$.
More precisely, we seek the solutions to the constrained optimization problem,
\begin{subequations}
\begin{equation}
\sup_{\vc u_0\in X} \big(I(\vc u(t_0+\tau))-I(\vc u(t_0))\big),
\label{eq:FTopt_sup}
\end{equation}
\begin{equation}
\mbox{where}\ 
\begin{cases}
\vc u(t)\ \mbox{satisfies}\ \eqref{eq:pde},\\
\vc C(\vc u_0)=\vc c_0,
\end{cases}
\label{eq:FTopt_const}
\end{equation}
\label{eq:FTopt}%
\end{subequations}
where the optimization variable is the initial condition $\vc u(t_0)=\vc u_0$
of system~(\ref{eq:pde}). 
The set of critical states are required to satisfy the constraints in~\eqref{eq:FTopt_const} in order to 
enforce two important properties. The first property ensures that $\vc u(t)$
obeys the governing equation~\eqref{eq:pde} as opposed to being an arbitrary one-parameter 
family of functions. The second property $\vc C(\vc u_0)=\mathbf c_0$, 
where $\vc C:X\to \mathbb R^k$, is a codimension-$k$ constraint. 
This constraint is enforced to ensure the non-zero probability of 
occurrence, i.e. states that are sufficiently close to the chaotic background attractor. 
The set of probabilistically feasible states 
can be generally described by exploiting basic physical properties of the chaotic attractor
such as average energy along different components of the state space or the second-order statistics. 
The precise form of the constraint $\vc C(\vc u_0)=\vc c_0$ 
is problem dependent and will shortly be discussed in more detail.
We point out that more general inequality constraints of the form 
$\vc c_{\mathrm{min}}\leq \vc C(\vc u_0)\leq \vc c_{\mathrm{max}}$ may also be employed. 
The treatment of such inequality constraints, however, is not discussed in this paper.

We expect the set of solutions to problem~\eqref{eq:FTopt} to unravel the mechanisms underpinning 
the intermittent bursts of the observable. Although it is unlikely that a generic trajectory of the system
passes exactly through one of the maximizers, by continuity, any trajectory passing through a sufficiently small open neighborhood of the maximizer (i.e., the instability regions of figure~\ref{fig:schem_FTgrowth}b) will 
result in a similar observable burst.

We emphasize that an optimization problem similar to~\eqref{eq:FTopt} has been pursued before in 
special contexts. The largest finite-time Lyapunov exponent can be formulated as~\eqref{eq:FTopt}
where the observable is the amplitude of infinitesimal perturbations after finite time. The 
maximizer is the corresponding finite-time Lyapunov vector~\cite{farrell96-1,farrell96-2}.
In a similar context, Pringle and Kerswell~\cite{pringle2010} seek optimal finite-amplitude perturbations that trigger transition to
turbulence in the pipe flow. They formulate the unknown optimal perturbation as 
the solution of a constrained optimization problem similar to~\eqref{eq:FTopt},
with the observable being the $L^2$ norm of the fluid velocity field.
Ayala and Protas~\cite{ayala2011,ayala2014,ayala2016} consider the 
finite-time singularity formation for Navier--Stokes equations. They also use a
variational method to seek the initial conditions that could lead to finite-time singularities. 
In these studies, the emphasis is given to the analysis of 
the most `unstable' states but the physical properties of the attractor are not taken into account.

The standard approach for solving the PDE-constrained optimization problem~(\ref{eq:FTopt}) is
an adjoint-based gradient iterative method~\cite{gunzburger2003,bartek04,faraz_cont}. 
This method is computationally very expensive
since, at each iteration, the gradient direction needs to be evaluated
as the solution of an adjoint PDE. If the growth timescale $\tau$ is small compared to the typical
time scales of the observable, it is reasonable to replace
the finite-time growth problem (\ref{eq:FTopt}) with its instantaneous 
counterpart,
\begin{subequations}
\begin{equation}
\sup_{\vc u_0\in X} \frac{\id}{\id t}\Big|_{t=t_0} I\big(\vc u(t)\big),
\label{eq:opt_ins_func}%
\end{equation}
\begin{equation}
\mbox{where}\ 
\begin{cases}
\vc u(t)\ \mbox{satisfies}\ \eqref{eq:pde},\\
\vc C(\vc u_0)=\vc c_0.
\end{cases}
\label{eq:opt_ins_const}%
\end{equation}
\label{eq:opt_ins}%
\end{subequations}
Problem~\eqref{eq:opt_ins} seeks initial states $\vc u_0$ for which the instantaneous growth of the observable
$I$ along the corresponding solution $\vc u(t)$ is maximal.

We point out that the large instantaneous derivatives of $I$ do not necessarily imply a subsequent burst in the observable as 
the growth may not always be sustained at later times along the trajectory $\vc u(t)$. 
As a result, the set of solutions to this instantaneous problem may differ 
significantly from the finite-time problem~\eqref{eq:FTopt}. 
Nonetheless, the solutions to the instantaneous problem can still be insightful. 
In addition, as we show below, these solutions can be obtained at a relatively low computational cost.

\subsection*{Optimal solutions}
First, we derive an equivalent form of problem~(\ref{eq:opt_ins}). Taking
the time derivative of the time series $I(\vc u(t))$ yields
$(\id/\id t) I(\vc u(t))=\id I(\vc u;\partial_t\vc u)$, where
$\id I(\vc u;\vc v):=\lim_{\varepsilon\to 0}\left[I(\vc u+\varepsilon\vc v)-I(\vc u)\right]/\varepsilon$ 
denotes the G\^ateaux differential of $I$ at 
$\vc u$ evaluated along $\vc v$. Using~\eqref{eq:pde_1} to substitute for
$\partial_t\vc u$, we obtain the following optimization problem
which is equivalent to problem~(\ref{eq:opt_ins}):
\begin{subequations}
\begin{equation}
\sup_{\vc u\in X} J(\vc u),
\end{equation}
\begin{equation}
\mbox{subject to}\ 
\begin{cases}
\mathcal K(\vc u)=0,\\
\vc C(\vc u)=\vc c_0,
\end{cases}
\end{equation}
\label{eq:opt_J}%
\end{subequations}
where 
\begin{equation}
J(\vc u):=\id I(\vc u;\mathcal N(\vc u)).
\label{eq:J}
\end{equation}
Note that the first constraint in~\eqref{eq:opt_ins_const}
simplified since we have already used~\eqref{eq:pde_1}
and it only remains to enforce~\eqref{eq:pde_2}.
For notational simplicity, we omit the subscript from $\vc u_0$.

If $J:X\to\mathbb R$ is a continuous map and the subset 
$S=\{\vc u\in X: \mathcal K(\vc u) =0,\; \vc C(\vc u)=\vc c_0\}$
is compact in $X$, problem (\ref{eq:opt_J}) has at least one solution.
This follows from the fact that the image of a compact set under
a continuous transformation is compact. Therefore $J(S)\subset\mathbb R$ is compact which implies 
$J(S)$ is bounded and closed~\cite{aliprantis98}. 
Therefore $J$ is bounded and attains its maximum (and minimum) on $S$.
The uniqueness of the maximizer is not generally guaranteed. 
However, the set of maximizers (and minimizers) of $J$
are compact subsets of $S$~\cite{aliprantis06}.

As we show in the Supplementary Material (section~\ref{app:proof_EL}), 
if $X$ is a Hilbert space with the inner product $\langle\cdot,\cdot\rangle_X$ and the operator $\mathcal K$ is linear, 
every solution of the optimization problem~\eqref{eq:opt_J} satisfies the set of Euler--Lagrange equations,
\begin{subequations}
\begin{equation}
J'(\vc u) + \mathcal K^\dagger(\alpha) +\sum_{i=1}^k\beta_i\, C_i'(\vc u)=\vc 0,
\label{eq:firstVari-1}
\end{equation}
\begin{equation}
\mathcal K(\vc u)=0,
\label{eq:firstVari-2}
\end{equation}
\begin{equation}
\vc C(\vc u)=\vc c_0.
\label{eq:firstVari-3}
\end{equation}
\label{eq:firstVari}%
\end{subequations}
Here $\mathcal K^\dagger$ is the adjoint of $\mathcal K$ and
$J'(\vc u)$ and $C_i'(\vc u)$ are the unique identifier of 
the G\^ateaux differentials $\id J(\vc u;\cdot)$ and $\id C_i(\vc u;\cdot)$, such that
$\id J(\vc u;\vc v)=\langle J'(\vc u),\vc v\rangle_X$ and 
$\id C_i(\vc u;\vc v)=\langle C_i'(\vc u),\vc v\rangle_X$
for all $\vc v$. The existence and uniqueness of 
$J'(\vc u)$ and $C_i'(\vc u)$ are guaranteed by the Riesz representation theorem~\cite{conway1985}.
Here, $C_i$ are the components of the map $\vc C=(C_1,C_2,\cdots,C_k)$.
The function $\alpha: \Omega\to \mathbb R$ and the vector 
$\vc \beta=(\beta_1,\cdots,\beta_k)\in\mathbb R^k$ are unknown Lagrange multipliers to be determined
together with the optimal state $\vc u:\Omega\to\mathbb R^d$.

\subsection*{Application to Navier--Stokes equations}
We consider the Navier--Stokes equations
\begin{equation}
\partial_t\vc u =-\vc u\cdot \bnabla\vc u -\bnabla p +\nu \Delta\vc u +\vc f,
\quad \bnabla\cdot\vc u =0,
\label{eq:nse_11}
\end{equation}
where $\vc u:\Omega\times\mathbb R^+\to \mathbb R^d$ is the fluid
velocity field, $p:\Omega\times\mathbb R^+\to\mathbb R$ is the
pressure field and $\nu=Re^{-1}$ is the non-dimensional viscosity
which coincides with the reciprocal of the Reynolds number $Re$.
Here, we consider two-dimensional flows ($d=2$) over the 
domain $\Omega=[0,2\pi]\times [0,2\pi]$ with periodic
boundary conditions. The flow is driven by the monochromatic 
Kolmogorov forcing $\vc f(\vc x) = \sin(k_f y)\vc e_1$ 
where $\vc k_f=(0,k_f)$ is the forcing wave number and
the vectors $\vc e_i$ denote the standard basis in $\mathbb R^d$.
In the following, we assume that the velocity fields are square integrable for all times, i.e., $X=L^2(\Omega)$.

The kinetic energy $E$ (per unit volume), the energy dissipation rate $D$ 
and the energy input rate $I$ are defined, respectively, by
\begin{equation}
E(\vc u)=\frac{1}{|\Omega|}\int_{\Omega} \frac{|\vc u|^2}{2}\id \vc x, \quad 
D(\vc u)=\frac{\nu}{|\Omega|}\int_{\Omega} |\bnabla\vc u|^2\id \vc x, \quad 
I(\vc u)=\frac{1}{|\Omega|}\int_{\Omega} \vc u\cdot \vc f\,\id \vc x,
\label{eq:IDE}
\end{equation}
where $|\Omega|$ denotes the area of the domain, i.e., $|\Omega|=(2\pi)^2$.
Along any trajectory $\vc u(t)$ these three quantities satisfy $\dot E = I-D$.
We use the energy dissipation rate $D$ to define the eddy turn-over time, 
$t_e=\sqrt{\nu/\mathbb E[D]}$, where $\mathbb E$ denotes the expected value.

The Kolmogorov flow admits the laminar solution $\vc u=(Re/k_f^2)\sin(k_fy)\vc e_1$.
For the forcing wave number $k_f=1$, the laminar solution is the global attractor of the system 
at any Reynolds number~\cite{marchioro86}. If the forcing is applied at a higher wave number
and the Reynolds number is sufficiently large, the laminar solution
becomes unstable.
In particular, numerical evidence suggests that for $k_f=4$ and sufficiently large Reynolds numbers, 
the Kolmogorov flow is chaotic and exhibits intermittent bursts of energy dissipation~\cite{PlSiFi91,CK13,faraz_adjoint}.
This is manifested in figure~\ref{fig:EngDiss}(a), showing the time series of the energy dissipation $D$ at Reynolds number $Re=40$
with $k_f=4$.
\begin{figure}
\centering
\includegraphics[width=.7\textwidth]{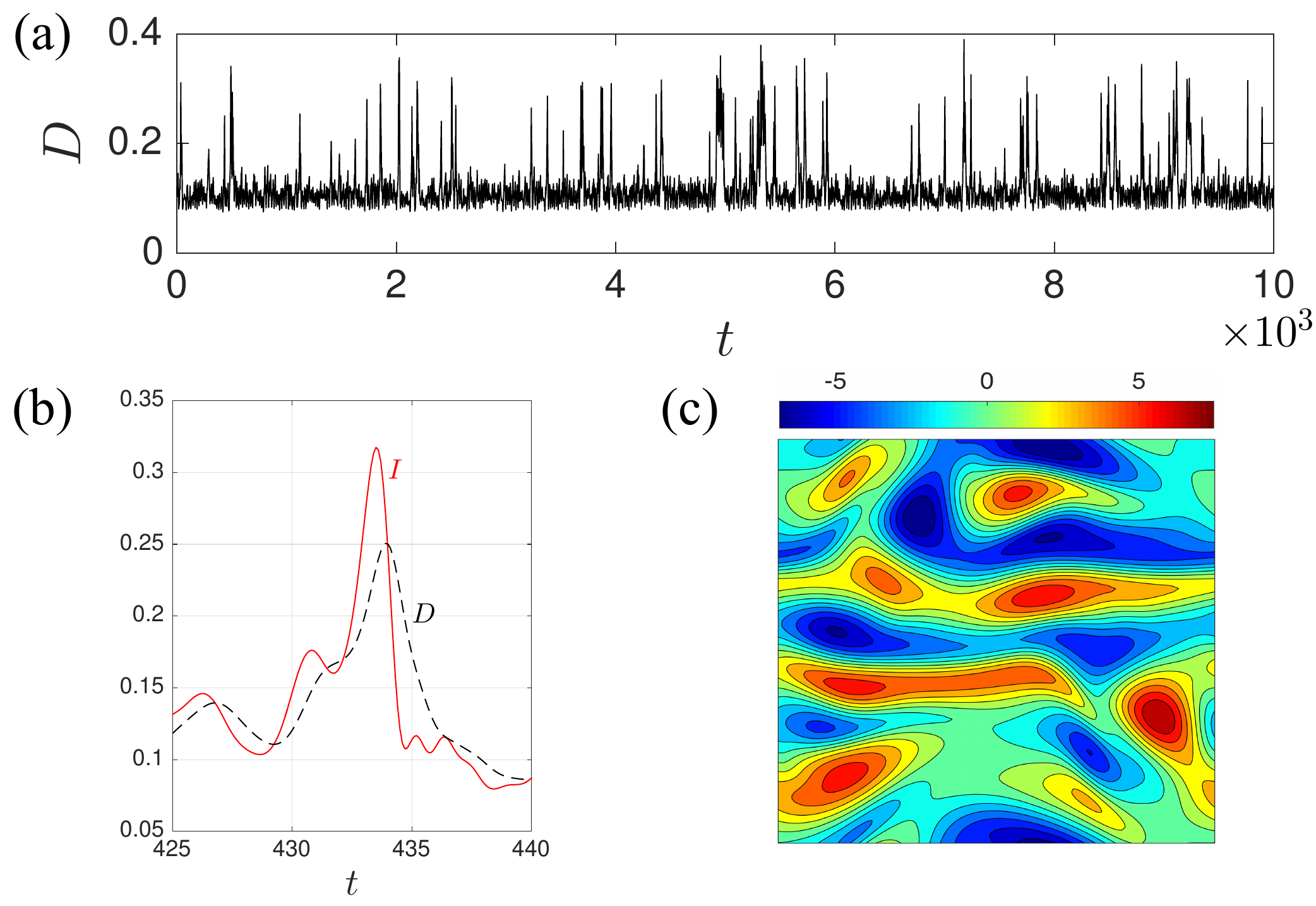}
\caption{(a) The time series of energy dissipation rate $D$ at Reynolds number $R=40$.
(b) A closeup of the energy input $I$ (solid red curves) and the energy dissipation $D$ (dashed black curves)
at $Re=40$. The bursts in the energy dissipation are slightly preceded with a burst in the energy input.
A similar behavior is observed for all bursts and at higher Reynolds numbers.
(c) The vorticity field $\bnabla\times \vc u(\vc x,t)=\omega(\vc x,t)\vc e_3$
 at time $t=433$ over the domain $\vc x\in[0,2\pi]\times[0,2\pi]$. 
}
\label{fig:EngDiss}
\end{figure}

A closer inspection reveals that each burst of the energy dissipation $D$ is shortly
preceded by a burst in the energy input $I$ (see figure~\ref{fig:EngDiss}(b)). Therefore, we 
expect the mechanism behind the bursts in the energy input to be also
responsible for the bursts in the energy dissipation. 
As we show in Supplementary Material (section~\ref{app:SI_nse_prelim}), the energy input is
given by $I(\vc u)=-|a(\vc k_f)|\sin(\phi(\vc k_f))$ where 
$a(\vc k)$ are the Fourier modes of the velocity field $\vc u$
and $\phi(\vc k)$ are their corresponding phases such that 
$a(\vc k)=|a(\vc k)|\exp(i\phi(\vc k))$.
We refer to the Fourier mode $a(\vc k_f)$ as the mean flow.
The energy input $I$ can grow through two mechanisms: 
(i) Alignment between the phase of the mean flow
and the external forcing, i.e., $\phi(\vc k_f)\to -\pi/2$ and 
(ii) Growth of the mean flow energy $|a(\vc k_f)|$.

Examining the alignment between the forcing and the velocity field rules out
mechanism (i) (cf. figure~\ref{fig:mean_phase_R40} of the Supplementary Material). 
The remaining mechanism (ii) is possible 
through the nonlinear term in the Navier--Stokes equation.
This nonlinearity redistributes the system energy among various
Fourier modes $a(\vc k)$ through triad interactions of the modes 
whose wave numbers $(\vc k,\vc p,\vc q)$ satisfy $\vc k=\vc p+\vc q$
(see Supplementary Material, section~\ref{app:dot_a}, for further details).
Due to the high number of active modes involved in the intricate
network of triad interactions, it is unclear which triad (or triads) are 
responsible for the nonlinear transfer of energy to the mean flow during
the extreme events. As we show below, our variational approach identifies the modes involved in
this transfer. Before obtaining the optimal solution, however, we need to 
specify the explicit form of the constraint $C(\vc u)=c_0$.

\subsubsection*{Constraints}\label{sec:const}
The constraint $\vc C(\vc u)=\vc c_0$ is imposed in order to ensure that
the optimal solutions are physically admissible, i.e. that they are sufficiently close to the attractor and thus have non-zero probability of occurrence. 
For instance, the solutions to a wide range of dissipative PDEs are known to converge asymptotically 
to a finite-dimensional subset of the state space~\cite{constantin2012}. 
The maximizers of the functionals~(\ref{eq:FTopt_sup}) 
or (\ref{eq:opt_ins_func}) that are far from this asymptotic attractor are 
physically irrelevant as they correspond to a transient phase that cannot be sustained. 

In most applications, including the Kolmogorov flow, the system attractor is not explicitly known. 
Therefore, the physical relevance of the optimal solutions need to be ensured
otherwise. Here, we consider constraints of the form 
\begin{equation}
C(\vc u):=\frac{1}{|\Omega|}\int_{\Omega}\frac{|A(\vc u)|^2}{2}\id\vc x,
\label{eq:const_A}
\end{equation}
where $A$ is a linear operator. Several physically important quantities 
can be written as the function (\ref{eq:const_A}). For instance, 
if $A$ is the identity operator, $C$ coincides with the kinetic energy, $C(\vc u)=E(\vc u)$.
If $A$ is the gradient operator, $A=\bnabla$, we have $C(\vc u)=D(\vc u)\times (Re/2)$ which can be used to
constrain the energy dissipation rate.
A more general class of such operators can be constructed as follows.
Let $\vc u =\sum_i \alpha_i\vc v_i$ where $\{\vc v_i\}$ is a principal component basis, i.e. it diagonalizes the covariance operator of $\vc u$.
Define $A$ as the diagonal linear operator such that $A(\vc v_i)=\vc v_i/\sigma_i$, where $\sigma_i^2$ is the standard deviation of $\alpha_i$. Then 
the constraint takes the form $C(\vc u)=\left( \sum_i \alpha_i^2/\sigma_i^2\right)/(2|\Omega|)$. This corresponds to an ellipsoid, describing points that have equal probability of occurrence when we approximate the statistics of the background attractor by a Gaussian measure~\cite{lumley1967}. 
Note that the constraints of the form~\eqref{eq:const_A} are
codimension-one, $C:X\to \mathbb R$, and hence a special case of the codimension-$k$
constraint in equation~\eqref{eq:FTopt_const}.

Excluding the intermittent bursts, the energy dissipation of the Kolmogorov 
flow exhibits small oscillations around its mean value $\mathbb E[D]$ (see figure~\ref{fig:EngDiss}a).
Based on this observation, we seek optimal solutions of~\eqref{eq:opt_J} which are constrained to have
the energy dissipation $D=\mathbb E[D]$. This result into the constraint (\ref{eq:const_A}) with $A=\bnabla$
and $C(\vc u)=c_0=\mathbb E[D]\times (Re/2)$. We approximate the mean value $\mathbb E[D]$ from
direct numerical simulations. At $Re=40$, for instance, we have $\mathbb E[D]\simeq 0.117$.

\subsubsection*{Probing the extreme energy transfers}
The functional $J$ (see Eq.~\eqref{eq:J}) associated with the energy input $I$ reads 
\begin{equation}
J(\vc u)=\frac{1}{|\Omega|}\int_{\Omega}\big[\vc u\cdot(\vc u\cdot\bnabla\vc f) +\nu\,\vc u\cdot\left(\Delta \vc f\right)\big] \id\vc x.
\end{equation}
The associated Euler--Lagrange equations (\ref{eq:firstVari}) read
\begin{subequations}
\begin{equation}
\left( \bnabla\vc f + \bnabla\vc f^\top \right)\vc u +\nu \Delta\vc f -\bnabla \alpha +\beta A^\dagger A\vc u=\vc 0,
\label{eq:uAlphaBetaPDE}
\end{equation}
\begin{equation}
\bnabla \cdot\vc u =0,
\label{eq:divFree}
\end{equation}
\begin{equation}
\frac{1}{|\Omega|}\int_{\Omega}\frac{|A(\vc u)|^2}{2}\id\vc x=c_0,
\label{eq:fixedE}
\end{equation}
\label{eq:opt_pde}%
\end{subequations}
where 
$J'(\vc u)=\left( \bnabla\vc f + \bnabla\vc f^\top \right)\vc u +\nu \Delta\vc f$, 
$\mathcal K^\dagger(\alpha) = -\bnabla\alpha$ and
$C'(\vc u)=A^\dagger A\vc u$ (see Supplementary Material, section~\ref{app:EL_NS}, for the derivations).
We set $A=\bnabla$ in order to enforce a constant energy dissipation constraint.
This implies $A^\dagger A\vc u = -\Delta \vc u$.
\begin{figure}
\centering
\includegraphics[width=.7\textwidth]{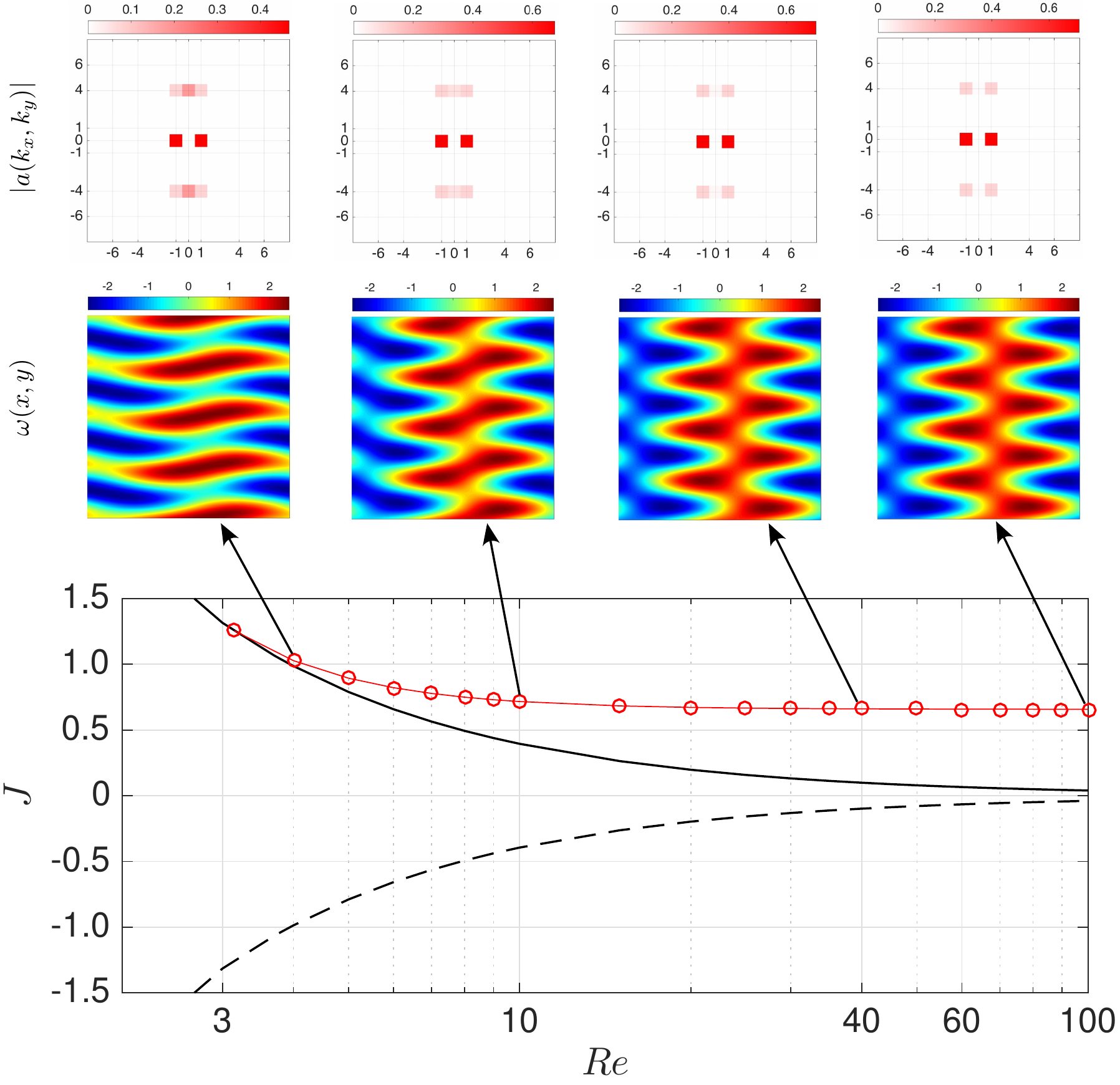}
\caption{The solutions of~\eqref{eq:opt_pde}, with $c_0=1$, as a function of the Reynolds number.
Solid (resp. dashed) black line corresponds to the exact solution $\vc u_{-}$ (resp. $\vc u_{+}$).
The red solid line (circles) corresponds to the global maximizer.
The outset shows the scalar vorticity $\bnabla\times \vc u(x,y)=\omega(x,y)\vc e_3$
and the Fourier spectrum $|a(k_x,k_y)|$
of the global maximizer at select Reynolds numbers.
}
\label{fig:optU}
\end{figure}

Using the symmetries of Eq.~\eqref{eq:opt_pde}, we find that it admits the pair of exact solutions
$\vc u_{\pm} =\pm(2\sqrt{c_0}/k_f)\sin(k_fy)\vc e_1$, 
$\alpha_{\pm}=\pm\sqrt{c_0}\int \sin(2k_fy)\id y$,
$\beta_{\pm}=\pm(\nu k_f/2\sqrt{c_0})$.
More complex solutions, with unknown closed forms, may exist. 
We approximate these solutions using the Newton iterations 
described in Supplementary Material, section~\ref{app:newton}. 

At each $Re$, we initiated several Newton iterations from
random initial conditions. In addition to the pair of exact solutions $(\vc u_\pm,\alpha_\pm,\beta_\pm)$,
the iterations yielded one non-trivial solution. 
Figure~\ref{fig:optU} shows the resulting three branches of solutions including
the exact solution $\vc u_+$ (solid black), the exact solution $\vc u_-$ (dashed black) and the
non-trivial solution (red circles). 
For small Reynolds numbers, our Newton searches
only returned the exact solutions. At $Re\simeq 3.1$,
a bifurcation takes place where a new non-trivial solution is born. This solution
appears to be a global maximizer as no other solutions were found. 
Since the intermittent bursts are only observed for $Re>35$, we focus 
the following analysis on this range of Reynolds numbers.
\begin{figure}
\centering
\includegraphics[width=.7\textwidth]{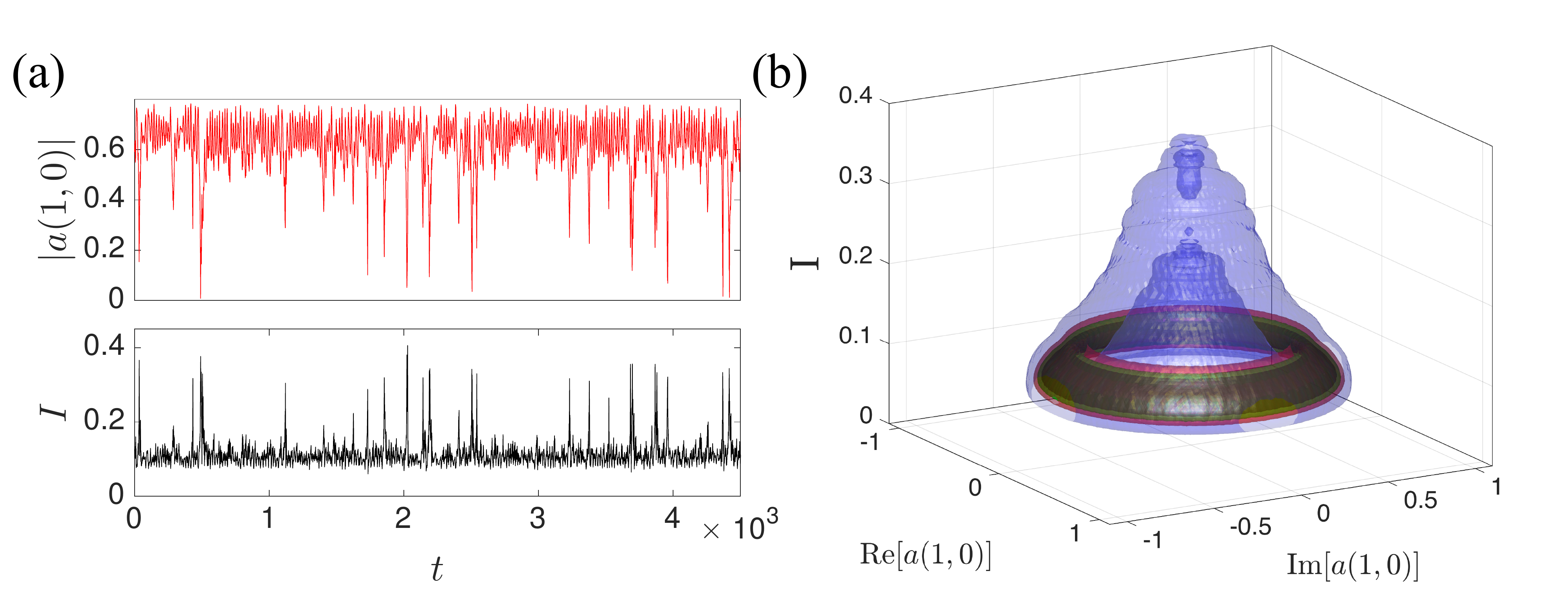}
\caption{(a) Time series of the energy input $I$ and the modulus of the Fourier mode $a(1,0)$ at $Re=40$.
The eddy turn-over time at this Reynolds number is $t_e=0.46$.
(b) The joint probability density of the energy input $I$ and the real and imaginary parts of 
the mode $a(1,0)$, approximated from $100,000$ samples. The density decreases from dark green to light blue.
The cone-shaped density indicates the strong correlation between the large values of 
the energy input rate $I$ and small values of $|a(1,0)|$.
The axisymmetric nature of the probability density is a consequence of the 
translation invariance of the Kolmogorov flow.}
\label{fig:I_vs_a10}
\end{figure}

The non-trivial optimal solution converges to an asymptotic 
limit as the $Re$ increases. 
This is discernible from the plateau of the red curve in figure~\ref{fig:optU}
and the select solutions shown in its outset. The three most dominant 
Fourier modes present in this asymptotic solution are the forcing
wave number $(0,k_f)$ and the wave numbers $(1,0)$ and $(1,k_f)$
together with their complex conjugate pairs.
Incidentally, these wave numbers form a triad, $(0,k_f)+(1,0)=(1,k_f)$.
The dominant mode of the optimal solution corresponds to the wave number $(1,0)$
whose modulus $|a(1,0)|$ is one order of magnitude larger than the other non-zero modes.

Next, we turn to the direct numerical simulations of the Kolmogorov flow
and monitor the three Fourier modes $a(0,k_f)$, $a(1,0)$ and $a(1,k_f)$.
We find that the energy transfers within this triad underpin the
intermittent bursts of the mean flow energy $|a(0,k_f)|$, and hence the energy input rate $I$.
Figure~\ref{fig:I_vs_a10}(a) shows time series of $I=-\im [a(0,k_f)]$ and the Fourier mode $|a(1,0)|$
along a typical trajectory of the Kolmogorov flow at $Re=40$. The bursts of the energy input rate $I$ are 
nearly concurrent with extreme dips in the modulus $|a(1,0)|$. A similar concurrent
behavior was observed for other trajectories and at 
higher Reynolds numbers (see Supplementary Material, section~\ref{app:results_highRe}).

This observation reveals that, prior to a burst, the mode $a(1,0)$
transfers a significant portion of its energy budget to the mean flow $a(0,k_f)$
through the triad interaction of the modes $a(0,k_f)$, $a(1,0)$ and $a(1,k_f)$.
This leads to the increase in the energy of the mean flow, and therefore
the energy input rate $I$, which in turn leads to the growth of the 
energy dissipation rate $D$.

One can go one step further and inquire about the reason for the release of energy from 
mode $a(1,0)$ to the mean flow $a(0,k_f)$. The answer to this question, 
involving the relative phases of the the modes and their interactions with
other triads, is beyond the scope of the present work and will be addressed elsewhere. 
It is tempting to study these interactions by truncating the Kolmogorov flow to
the modes $a(0,k_f)$, $a(1,0)$, $a(1,k_f)$ and their complex conjugates. Unfortunately, 
such severe truncations fail to be illuminating since the dynamics of
the truncated system severely departs from the original Navier--Stokes equations~\cite{moffatt2014,Biferale16}.

\subsubsection*{Prediction of extreme events}\label{sec:Pee}
Given the above observation that the optimal solution primarily
consists of mode $(1,0)$, we choose this mode to formulate our data-driven prediction scheme.
The decrease in the energy of the mode $(1,0)$ precedes the 
increase in the energy of the mean flow. This enables the data-driven short-term
prediction of extreme bursts of the energy dissipation by observing the 
modulus $|a(1,0,t)|$. More specifically,
relatively small values of $\lambda(t):=|a(1,0,t)|$, along a solution $\vc u(t)$, signal
the high probability of an upcoming burst in the energy dissipation. 
\begin{figure}
\centering
\includegraphics[width=.7\textwidth]{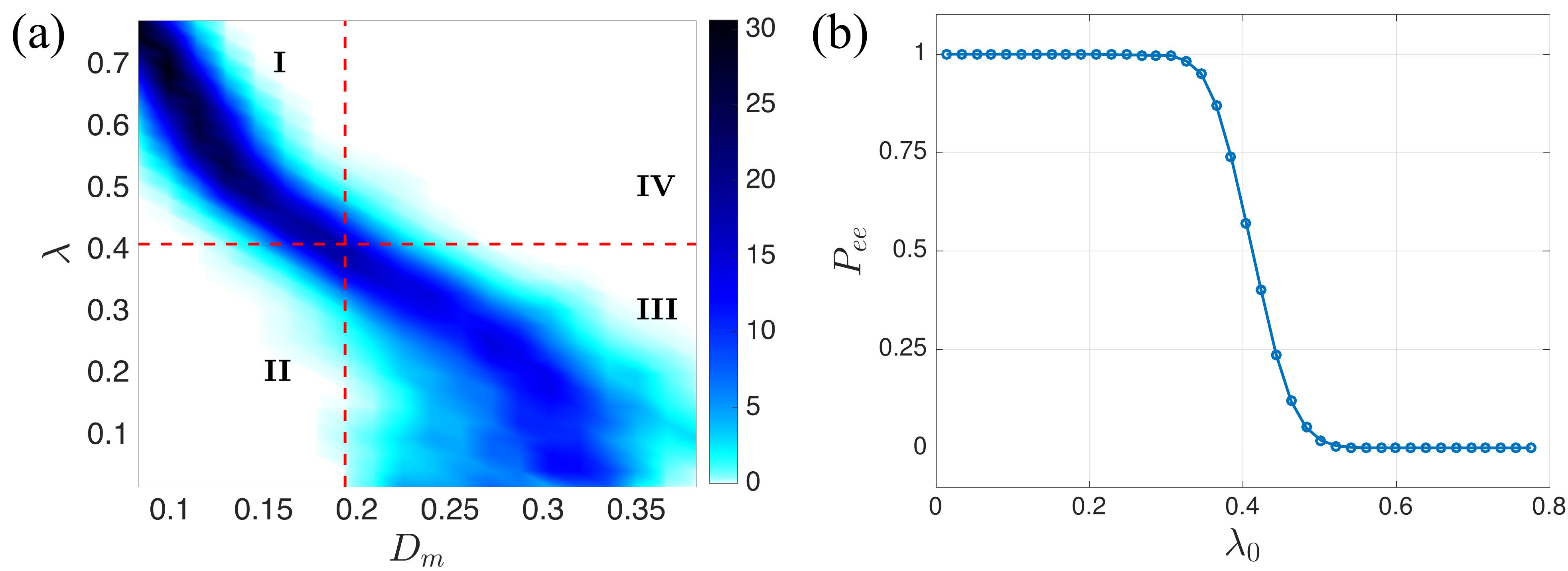}
\caption{
(a) The probability density assocaited with the conditional probability (\ref{eq:condPDF_timeShift}).
The vertical dashed line marks the extreme event threshold 
$D_e=\mathbb E[D]+2\sqrt{\mathbb E[D^2]-\mathbb E[D]^2}\simeq 0.194$.
The horizontal dashed line marks $\lambda=0.4$.
The quadrants correspond to 
I. Correct rejections, 
II. False positives, 
III. Hits and 
IV. False negatives.
(b) The probability of extreme events $P_{ee}$
corresponding to the extreme event threshold $D_e= 0.194$.
}
\label{fig:condPDF_timeShift}
\end{figure}

To quantify this, we consider the conditional probability
\begin{align}
\mathcal P\left(D_1\leq \Dmax(t) \leq D_2 \big|\, \lambda_1\leq \lambda(t)\leq \lambda_2\right),
\label{eq:condPDF_timeShift}
\end{align}
where 
$\Dmax(t)=\max_{\tau\in[t+t_i,t+t_f]} D(\vc u(\tau))$
is 
the maximum of the energy dissipation
over the future time interval $[t+t_i,t+t_f]$.
This conditional probability measures the likelihood of the future maximum value of the
energy dissipation belonging to the interval 
$[D_1,D_2]$, given that the present value of the indicator $\lambda(t)$
belongs to $[\lambda_1,\lambda_2]$. The constant parameters $t_f>t_i>0$ 
determine the future time interval $[t+t_i,t+t_f]$. In the following, 
we set $t_i=1\simeq 2.2t_e$ and $t_f=2\simeq t_i+2.2t_e$. The length of the time window
$t_f-t_i$ is long enough to ensure that the extreme event (if it exists) is contained in the
time interval $[t+t_i,t+t_f]$. The choice of the prediction time $t_i$ will be discussed shortly.
The reported results are robust to small perturbations to all parameters. 

Figure~\ref{fig:condPDF_timeShift}(a)
shows the conditional probability density corresponding to~\eqref{eq:condPDF_timeShift}.
We observe that relatively small values of $\lambda$ correlate strongly with the 
high future values of the energy dissipation $D$. For instance,
when $\lambda<0.4$, the value of $D_m$ is most likely larger than $0.2$.
Conversely, when $\lambda$ is larger than $0.4$, 
the future values of the future energy dissipation $D_m$ are smaller than $0.2$.

We seek an appropriate value $\lambda_0$
such that $\lambda(t)<\lambda_0$ predicts an extreme burst of energy dissipation over the
future time interval $[t+t_i,t+t_f]$.
Denote the extreme event threshold by $D_e$, such that $D>D_e$
constitutes an extreme burst of energy dissipation.
We define the probability of an upcoming extreme event $P_{ee}$ as
\begin{equation}
P_{ee}\left(\lambda_0\right) = \mathcal P\left( \Dmax(t)>D_e \big|\, \lambda(t)=\lambda_0\right),
\label{eq:Pee}
\end{equation}
which measures the likelihood that $D_m(t)>D_e$ assuming that $\lambda(t)=\lambda_0$.
Here, we set the threshold of the extreme event $D_e$ as the mean value of the energy dissipation plus two
standard deviations, $D_e=\mathbb E[D]+2\sqrt{\mathbb E[D^2]-\mathbb E[D]^2}\simeq 0.194$.
The extreme event probability $P_{ee}$ can be computed from the probability density shown in figure~\ref{fig:condPDF_timeShift}(a) (see Supplementary Material, section~\ref{app:Pee}, for the details).

Figure~\ref{fig:condPDF_timeShift}(b) shows the probability of extreme events $P_{ee}$ 
as a function of the parameter $\lambda_0$. If at time $t$, the values of $\lambda(t)$ is larger
than $0.5$ the probability of a future extreme event, i.e. $D_m(t)>D_e$, is nearly zero. 
The probability of a future extreme event increases as $\lambda(t)$ decreases.
At $\lambda(t)\simeq 0.4$, the probability is $50\%$. If $\lambda(t)< 0.3$,
the likelihood of an upcoming extreme event is nearly $100\%$.
The horizontal dashed line in figure~\ref{fig:condPDF_timeShift}(a)
marks the transition line from the low likelihood of an upcoming extreme event $P_{ee}<0.5$
to the higher likelihood $P_{ee}>0.5$. This line together with the vertical line $D_m=D_e$
divide the conditional probability density into four regions: 
(I) Correct rejections ($P_{ee}<0.5$ and $D_m(t)<D_e$): Correct prediction of no upcoming extremes.
(II) False positives ($P_{ee}>0.5$ but $D_e(t)<D_e$): The indicator predicts an upcoming extreme event 
but no extreme event actually takes place.
(III) Hits ($P_{ee}>0.5$ and $D_m(t)>D_e$): Correct prediction of an upcoming extreme event.
(IV) False negatives ($P_{ee}<0.5$ but $D_m(t)>D_e$): An extreme event takes place but
the indicator fails to predict it.

A reliable indicator of upcoming extreme events must
maximize the number of correct rejections (quadrant I)
and hits (quadrant III), while at the same time 
having minimal false positives (quadrant II) and false negatives (quadrant IV). 
From nearly $100,000$ predictions made only $0.26\%$ false negatives and
$0.85\%$ false positives were recorded. The number of hits were $5.6\%$ 
and the number of correct rejections amount to $93.3\%$ of all predictions made.
As we show in the Supplementary Material, this amounts to $95.6\%$ success rate for the prediction of the extreme events
(see Equation~\eqref{eq:rsp} and Table~\ref{tab:params}).
Note that the high percentage of correct rejections compared to the hits is a mere consequence of 
the fact that the extreme events are \emph{rare}. 

An additional desirable property of an indicator is its ability to 
predict the upcoming extremes well in advance of the
events taking place. The chosen prediction time $t_i=1$
is approximately twice the eddy turn-over time $t_e$. In comparison, 
it takes approximately one eddy turn-over time (on average) 
for the energy dissipation rate to grow from
$\mathbb E[D]+\sqrt{\mathbb E[D^2]-\mathbb E[D]^2}$
to its extreme value 
$D_e=\mathbb E[D]+2\sqrt{\mathbb E[D^2]-\mathbb E[D]^2}$.

The prediction time $t_i$ can always be increased at the cost of 
increasing false positives and/or false negatives. For instance, 
with the choice $t_i=2\simeq 4.3t_e$ and $t_f=3\simeq t_i+2.2t_e$,
prediction of the extreme events $D_m>D_e$ returns $1.2\%$ false negatives and $0.6\%$ 
false positives. The number of hits decreases slightly to $5.3\%$, as does the number of
correct rejections $92.9\%$, 
which amounts to a success rate of $82\%$ in the extreme event 
prediction (see Equation~\eqref{eq:rsp}). 
Therefore, the prediction time $t_i=2$ still yields satisfactory predictions.
Upon increasing $t_i$ further, eventually the number of hits becomes comparable to
the number of false negatives at which point the predictions
are unreliable.

\section*{Discussion}
A method for the computation of precursors of extreme events in complex turbulent systems is introduced here. The  new approach combines basic physical properties of the chaotic attractor (such as energy distribution along different directions of phase space) obtained from data, with stability properties induced by the governing equations. The method is formulated as a constrained optimization problem which can be solved explicitly if the timescale of the extreme events is short compared to the typical timescales of the system. To demonstrate the approach we consider a stringent test case, the Kolmogorov flow, which has a turbulent attractor with positive Lyapunov exponents and intermittent extreme bursts of energy dissipation. We are able to correctly identify the triad of modes associated with the extreme events. Moreover, the derived precursors allow for the formulation of an accurate short-term prediction scheme for the intermittent bursts. The results demonstrate the robustness and applicability of the approach on systems with high-dimensional chaotic attractors. 

\section*{Materials and Methods}
%

The Navier--Stokes equations and the corresponding Euler--Lagrange equations are solved numerically with a standard 
pseudo-spectral code with $N\times N$ Fourier modes and $2/3$ dealiasing. For $Re = 60, 80$ and $100$, we use
$N=256$ to fully resolve the velocity fields. At $Re = 40$, however, this resolution is unnecessarily high and hence we use 
$N=128$. The temporal integration of the Navier--Stokes
equations are carried out with a forth-order Runge--Kutta scheme.


\section*{Supplementary Material}
section S1. Derivation of the Euler--Lagrange equation\\
section S2. The Navier--Stokes equation\\
section S3. Newton iterations\\
section S4. Sensitivity to parameters\\
section S5. Computing the probability of extreme events\\
section S6. Supporting computational results\\
fig. S1. Evolution of the energy input vs. mean flow\\
fig. S2. Triad interactions \\
fig. S3. Sensitivity of the optimal solutions\\
fig. S4. Joint PDFs for higher Reynolds numbers\\
fig. S5. Prediction of intermittent bursts at higher Reynolds numbers\\
table S1: Simulation parameters

\noindent \textbf{Acknowledgments:} This work has been supported through the ONR grant N00014-15-1-2381, the AFOSR grant FA9550-16- 1-0231, and the ARO grant 66710-EG-YIP.\\
\noindent \textbf{Data and materials availability:}
Certain codes and data used in this manuscript are publicly available on GitHub at 
\href{https://github.com/mfarazmand/VariExtEvent}{https://github.com/mfarazmand/VariExtEvent}.

\clearpage
\renewcommand{\thesection}{S\arabic{section}}
\renewcommand{\thefigure}{S\arabic{figure}}
\renewcommand{\thetable}{S\arabic{table}}
\renewcommand{\theequation}{S\arabic{equation}}

\section*{Supplementary Material}



\section{Derivation of the Euler--Lagrange equation}\label{app:proof_EL}
In this section, we detail the derivation of the Euler--Lagrange equations.
We first form the constrained Lagrangian functional,
\begin{equation}
\mathcal L_c(\vc u,\alpha,\vc \beta):=J(\vc u)+\langle \mathcal K(\vc u),\alpha\rangle_X + 
\vc\beta\cdot (\vc C(\vc u)-\vc c_{0}),
\end{equation}
where the function $\alpha : \Omega\to\mathbb R$ and the vector $\vc\beta =(\beta_1,\cdots,\beta_k)\in\mathbb R^k$ 
are the Lagrange multipliers. Taking the first variation of the constrained Lagrangian with respect to $\vc u$, we obtain
\begin{align}
\frac{\delta\mathcal L_c}{\delta\vc u}(\vc v) & := \lim_{\varepsilon\to 0}\frac{1}{\varepsilon}\left[\mathcal L(\vc u+\varepsilon\vc v,\alpha,\vc\beta)- \mathcal L(\vc u,\alpha,\vc\beta)\right]\nonumber\\
& = \id J(\vc u;\vc v) + \langle\mathcal K(\vc v),\alpha\rangle +\sum_{i=1}^{k}\beta_i \id C_i(\vc u;\vc v)\nonumber\\
& = \langle J'(\vc u),\vc v\rangle + \langle\mathcal K^\dagger(\alpha),\vc v\rangle + \sum_{i=1}^{k}\beta_i \langle C'_i(\vc u),\vc v\rangle\nonumber\\
& = \langle J'(\vc u) + \mathcal K^\dagger(\alpha) + \edit{\sum_{i=1}^{k}\beta_i C'_i(\vc u)},\vc v\rangle.
\end{align}
Since the first variation $\delta\mathcal L_c/\delta \vc u$ must vanish for all $\vc v$, we obtain
\begin{equation}
 J'(\vc u) + \mathcal K^\dagger(\alpha) + \edit{\sum_{i=1}^{k}\beta_i C'_i(\vc u)}=\vc 0.
\label{eq:EL}
\end{equation}
Similarly, the first variations of the Lagrangian $\mathcal L_c$ with respect to the Lagrange multipliers $\alpha$ and $\vc\beta$ read
\begin{equation}
\frac{\delta\mathcal L_c}{\delta\alpha}(\tilde\alpha) = \langle\mathcal K(\vc u),\tilde{\alpha}\rangle_X,\quad 
\frac{\delta\mathcal L_c}{\delta\vc\beta} = \vc C(\vc u)-\vc c_0.
\label{eq:EL-const}
\end{equation}
Since they must vanish for all $\tilde{\alpha}$, we obtain the constraints $\mathcal K(\vc u)=0$ and
$\vc C(\vc u)=\vc c_0$.

\section{The Navier--Stokes equation}
\subsection{Preliminaries}\label{app:SI_nse_prelim}
Recall the Navier--Stokes equations
\begin{subequations}
\begin{equation}
\partial_t\vc u =-\vc u\cdot \bnabla\vc u -\bnabla p +\nu \Delta\vc u +\vc f,
\label{eq:nse_1}
\end{equation}
\begin{equation}
\bnabla\cdot\vc u =0,
\label{eq:nse_2}
\end{equation}
\label{eq:nse}%
\end{subequations}
with the Kolmogorov forcing $\vc f(\vc x) = \sin(k_f y)\vc e_1$ for some forcing wave number
$\vc k_f=(0,k_f)$. In two dimensions, a divergence free velocity field $\vc u:\Omega \to \mathbb R^2$ 
admits the following Fourier series expansion,
\begin{equation}
\vc u(\vc x,t)=\sum_{\vc k\in\mathbb Z^2} \frac{a(\vc k,t)}{k}
\begin{pmatrix}
k_2\\
-k_1
\end{pmatrix}e^{\hat i\vc k\cdot\vc x},
\label{eq:u_ft}
\end{equation}
where $\vc k=(k_1,k_2)$, $k=|\vc k|$ and $\hat i =\sqrt{-1}$ (see Ref.~\cite{PlSiFi91}). Since the 
velocity field is real-valued, we have $a(-\vc k)=-\overline{a(\vc k)}$.

For the Kolmogorov forcing, the energy input rate satisfies
\begin{equation}
I(\vc u(t))=-\im[a(\vc k_f,t)]=-r(\vc k_f,t)\sin\left(\phi(\vc k_f,t)\right),
\label{eq:I_a}
\end{equation}
where $\im$ denotes the imaginary part and $a(\vc k,t)=r(\vc k,t)\exp(\hat i\phi (\vc k,t))$ is the 
Fourier coefficient with phase $\phi(\vc k,t)\in(-\pi,\pi]$ and amplitude $r(\vc k,t)\in\mathbb R^+$. 
For simplicity, we may omit the dependence of these variables on time $t$.
For reasons that will become clear in the next section, we refer to the Fourier mode $a(\vc k_f,t)$
as the mean flow.

\begin{figure}
\centering
\includegraphics[width=.7\textwidth]{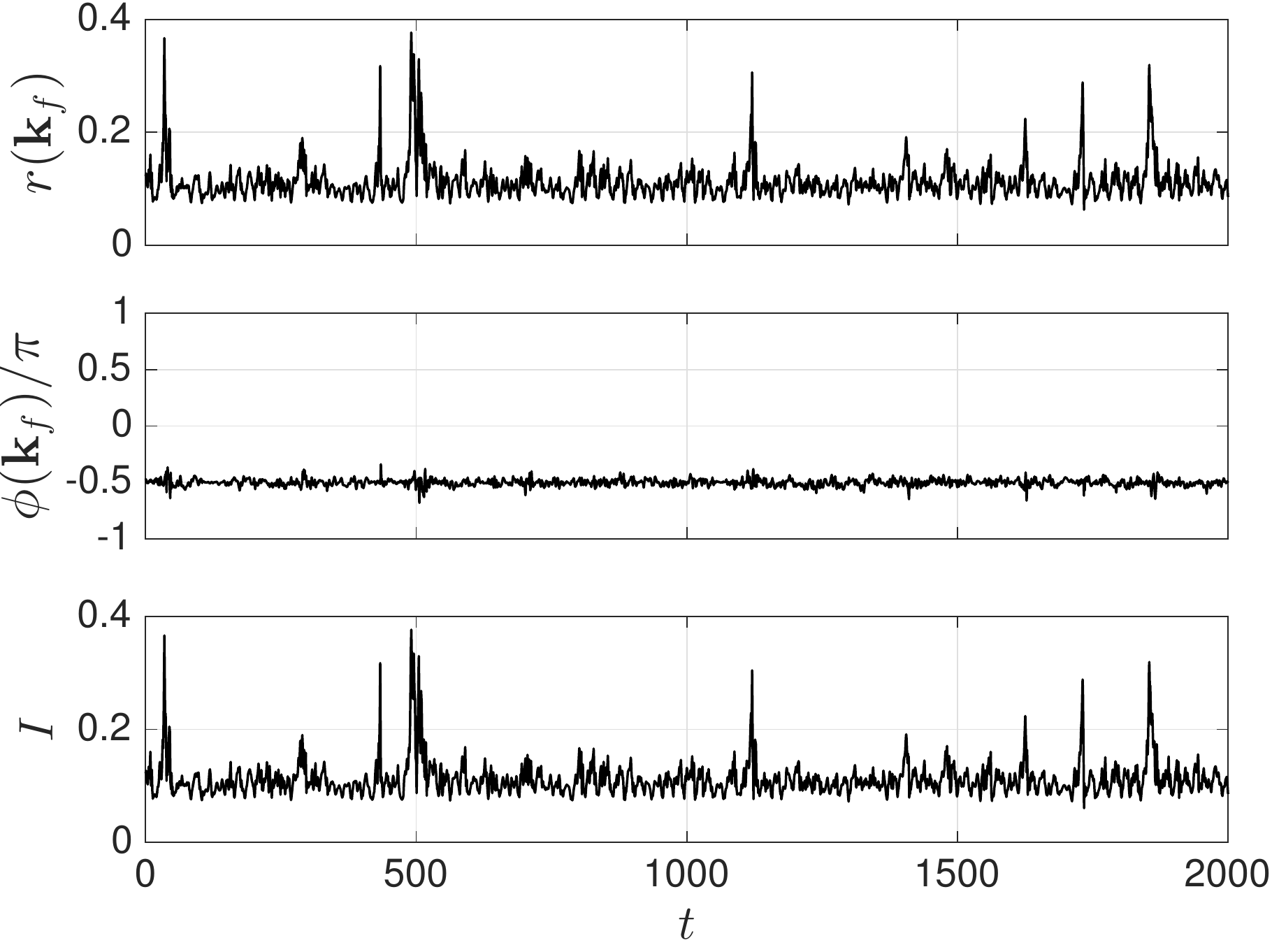}
\caption{The evolution of the energy input $I=-\im[a(\vc k_f)]=-r(\vc k_f)\sin [\phi(\vc k_f)]$, 
the phase $\phi(\vc k_f)$ of the mean flow and 
the amplitude $r(\vc k_f)$ of the mean flow along a typical trajectory of the Kolmogorov flow 
at $Re=40$. Note that the phase of the external force is $-\pi/2$. The forcing wave number is $\vc k_f=(0,4)$.}
\label{fig:mean_phase_R40}
\end{figure}

Examining equation~\eqref{eq:I_a}, the energy input $I$ may grow through two mechanisms: 
\begin{enumerate}[(1)]
\item The phase $\phi(\vc k_f)$ approaching $-\pi/2$,
\item The amplitude $r(\vc k_f)$ growing.
\end{enumerate}
Noting that the phase of the external forcing is also $-\pi/2$, 
scenario (1) corresponds to an alignment between the phases of
the external force and the mean flow $a(\vc k_f)$.
It is therefore tempting to attribute the intermittent bursts of the energy input $I$ 
to the intermittent alignments between the forcing $\vc f$
and the velocity field $\vc u$.  This postulate, however, does not stand further scrutiny. 
Figure~\ref{fig:mean_phase_R40}
shows the phase $\phi(\vc k_f,t)$ of the mean flow along a typical 
Kolmogorov trajectory $\vc u(t)$. This phase oscillates around $-\pi/2$
for all times. Note that $-\pi/2$ corresponds to perfect alignment 
between the mean flow and external forcing. Figure~\ref{fig:mean_phase_R40}
also shows the evolution of the energy input $I$ along the same trajectory.
No positive correlation exists between intermittent growth of the mean flow 
energy $I$ and the phase of the mean flow being $-\pi/2$. In fact, the 
phase $\phi(\vc k_f)$ seems to deviate from $-\pi/2$ during the bursts. 
Contrast this with the strong correlation between the growth of the
energy input rate and the amplitude $r(\vc k_f)$ of the mean flow.

This observation shows that the intermittent energy input bursts 
are triggered through mechanism (2), that is the growth of the amplitude $r(\vc k_f)$.
A similar observation is made at higher Reynolds numbers (not shown here).
This growth of the mean flow amplitude, in turn, is possible through the internal transfer of energy via nonlinear terms
as discussed below.

\subsection{Nonlinear triad interactions}\label{app:dot_a}
\begin{figure}
\centering
\includegraphics[width=.6\textwidth]{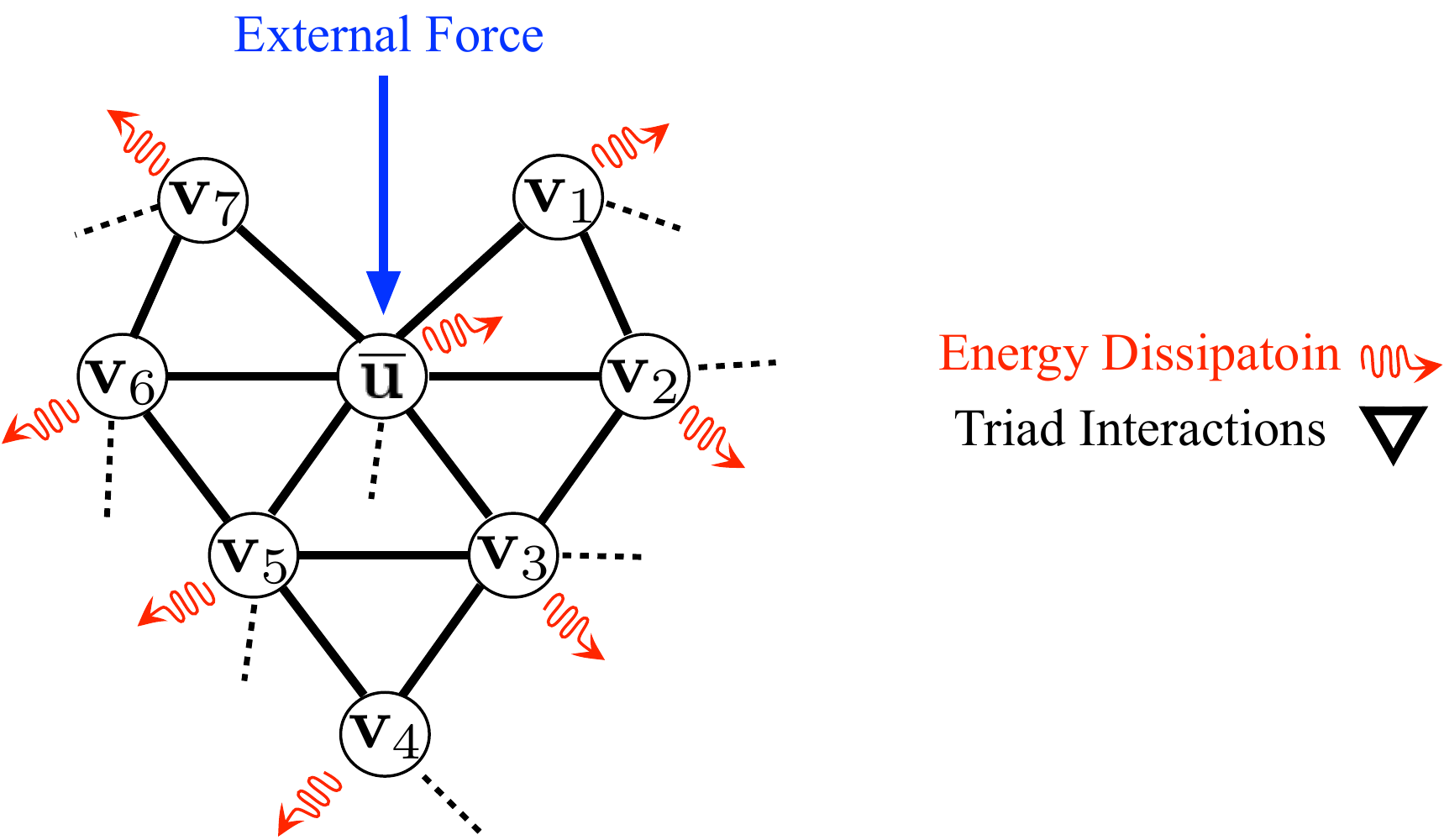}
\caption{Schematic representation of the triad interactions of the Navier--Stokes equation.}
\label{fig:schem_trid}
\end{figure}

The velocity field $\vc u(\vc x,t)$ can be written in the general Fourier-type expansion
\begin{equation}
\vc u(\vc x,t)=\overline{\vc u}(\vc x,t)+\sum_{j=1}^\infty \alpha_j(t)\vc v_j(\vc x),
\end{equation}
where $\overline{\vc u}$ is the statistical mean and $\{\vc v_j\}$ is a set of prescribed functions
that form a complete basis for the function space $X$ ($=L^2(\Omega)$). 
Under certain assumptions which are met by the Navier--Stokes equation~\eqref{eq:nse},
the energy is injected into the mean flow $\overline{\vc u}$ by the external forcing $\vc f$ (see Refs.~\cite{sapsis2013attractor,majda2015}).
The nonlinear term $\vc u\cdot\bnabla \vc u$, coupling the mean flow and the modes $\vc v_j$,
redistributes the injected energy to all modes $\vc v_j$. This nonlinear term conserves the total
energy of the system. At the same time, each mode dissipates energy due to the viscous term $\nu\Delta \vc u$
(see figure~\ref{fig:schem_trid}, for an illustration). A convenient choice of the basis $\{\vc v_j\}$ 
is problem dependent. Here, we choose the conventional Fourier basis as described in equation~\eqref{eq:u_ft}.
In case $\vc f$ is the Kolmogorov forcing, the symmetries of the system dictate 
$\overline{\vc u}(\vc x,t)=\alpha_0(t)\vc f(\vc x)=\alpha_0(t)\sin(k_fy)\vc e_1$ (see, e.g.,~\cite{CK13,faraz_adjoint}).

In order to make the above statements more explicit, we write the Navier--Stokes
equation in the Fourier space. Following~\cite{kraichnan1967}, we have
\begin{equation}
\partial_t \hat u_i(\vc k)=-\hat i P_{ij}(\vc k)\sum_{\vc p+\vc q=\vc k}q_m\hat u_m(\vc p)\hat u_j(\vc q)
-\nu k^2 \hat u_i(\vc k)+\hat f_i(\vc k),
\label{eq:nse_fourier1}
\end{equation}
where the hat signs denote the Fourier transform, $P_{ij}(\vc k)=\delta_{ij}-k_ik_j/k^2$
is the Leray projection onto the space of diverge-free vector fields
and the convention of summation over repeated indices is used.
Equation~\eqref{eq:nse_fourier1} can be written more explicitly as
\begin{subequations}
\begin{align}
\partial_t \hat u_1(\vc k)=&-\hat i\sum_{\vc p+\vc q=\vc k}q_m\hat u_m(\vc p)\left[\left(1-\frac{k_1^2}{k^2}\right)\hat u_1(\vc q)-\frac{k_1k_2}{k^2}\hat u_2(\vc q) \right]-\nu k^2 \hat u_1(\vc k)+\hat f_1(\vc k)
\end{align}
\begin{align}
\partial_t \hat u_2(\vc k)=&-\hat i\sum_{\vc p+\vc q=\vc k}q_m\hat u_m(\vc p)\left[-\frac{k_1k_2}{k^2}\hat u_1(\vc q)+\left(1-\frac{k_2^2}{k^2}\right)\hat u_2(\vc q) \right] -\nu k^2 \hat u_2(\vc k)+\hat f_2(\vc k)
\label{eq:nse_fourier2}
\end{align}
\end{subequations}

Recall the Fourier expansion~\eqref{eq:u_ft} which implies $\hat u_1(\vc k)=k_2 a(\vc k)/k$ and $\hat u_2(\vc k)=-k_1 a(\vc k)/k$.
Upon substitution in equation~\eqref{eq:nse_fourier2} and noting that
$$q_m\hat u_m(\vc p)=\frac{q_1p_2-q_2p_1}{p}a(\vc p),$$
and 
$$ \hat f_1(\vc k)=\frac{1}{2}e^{-\hat i\frac{\pi}{2}}\delta_{\vc k,\vc k_f}+\frac{1}{2}e^{+\hat i\frac{\pi}{2}}\delta_{\vc k,-\vc k_f},\quad
\hat f_2(\vc k)=0,$$
we obtain
\begin{equation}
\dot a(\vc k)=-\hat i \sum_{\vc p+\vc q=\vc k}\frac{(q_1p_2-q_2p_1)(k_1q_1+k_2q_2)}{p\,q\,k}a(\vc p)a(\vc q)
-\nu k^2a(\vc k)
+\frac{1}{2}e^{-\hat i\frac{\pi}{2}}\left(\delta_{\vc k,\vc k_f}+\delta_{\vc k,-\vc k_f}\right).
\end{equation}
We rewrite the above equation more compactly,
\begin{equation}
\dot a(\vc k)=\hat i \sum_{\vc p+\vc q=\vc k}\frac{\mu(\vc p,\vc q)(\vc k\cdot \vc q)}{p\,q\,k}a(\vc p)a(\vc q)
-\nu k^2a(\vc k)
+\frac{1}{2}e^{-\hat i\frac{\pi}{2}}\left(\delta_{\vc k,\vc k_f}+\delta_{\vc k,-\vc k_f}\right),
\label{eq:triadODE}
\end{equation}
where $\vc k\cdot \vc q = k_1q_1+k_2q_2$ and 
$\mu(\vc p,\vc q):=p_1q_2-p_2q_1$ is the two-form measuring the surface area of the parallelogram with sides $\vc p$ and $\vc q$.
Writing the modes in terms of their amplitudes and phases,  $a(\vc k)=r(\vc k)\exp[\hat i \phi(\vc k)]$, and using 
equation~\eqref{eq:triadODE}, we obtain
\begin{subequations}
\begin{align}
\dot r(\vc k)  = & 
\frac{1}{2}\cos\left[\frac{\pi}{2}+\phi(\vc k) \right]\left(\delta_{\vc k,\vc k_f}+\delta_{\vc k,-\vc k_f}\right)
-\nu k^2r(\vc k) \nonumber\\
& + \sum_{\vc p+\vc q=\vc k}\frac{\mu(\vc p,\vc q)(\vc k\cdot \vc q)}{p\,q\,k}
r(\vc p)r(\vc q)\sin\left[\phi(\vc k)-\phi(\vc p)-\phi(\vc q)\right],
\label{eq:nse_PhaseAmp_01}
\end{align}
\begin{align}
\dot \phi(\vc k)  = &
-\frac{1}{2}\frac{1}{r(\vc k)}\sin\left[\frac{\pi}{2}+\phi(\vc k) \right]\left(\delta_{\vc k,\vc k_f}+\delta_{\vc k,-\vc k_f}\right) \nonumber\\
&+ \sum_{\vc p+\vc q=\vc k}\frac{\mu(\vc p,\vc q)(\vc k\cdot \vc q)}{p\,q\,k}
\frac{r(\vc p)r(\vc q)}{r(\vc k)}\cos\left[\phi(\vc k)-\phi(\vc p)-\phi(\vc q)\right].
\label{eq:nse_PhaseAmp_02}
\end{align}
\label{eq:nse_PhaseAmp}
\end{subequations}
Note that $a(-\vc k)=-\overline{a(\vc k)}$ implies $\phi(-\vc k)=\pi - \phi(\vc k)$.

We now focus on the amplitude of the mean flow $r(\vc k_f)$
(and its corresponding conjugate at $\vc k=-\vc k_f$). The negative definite term $-\nu k_f^2 r(\vc k_f)$ representing the 
dissipation acts to decrease the mean flow amplitude.
This decay is counteracted by the external forcing $\frac{1}{2}\cos\left[\frac{\pi}{2}+\phi(\vc k_f) \right]$.
Recall from figure~\ref{fig:mean_phase_R40} that the phase $\phi(\vc k_f)$ oscillates around $-\pi/2$ 
for all times, implying $\cos\left[\frac{\pi}{2}+\phi(\vc k_f) \right]>0$.
The complications arise from the summation term in~\eqref{eq:nse_PhaseAmp_01}
which couples the mean flow to all other modes that form the wave vector triads, $\vc p+\vc q=\vc k_f$.
The contribution from these other modes depends on the amplitudes, $r(\vc p)$ and $r(\vc q)$, and the
relative phases $\phi(\vc k_f)-\phi(\vc p)-\phi(\vc q)$. Even the modes that do not form a 
triad with $\vc k_f$, affect the mean flow amplitude indirectly through their coupling to the modes that do
form a triad with $\vc k_f$ (see the schematic figure~\ref{fig:schem_trid}).

\subsection{Derivation of Euler-Lagrange equation for Navier--Stokes}\label{app:EL_NS}
We first derive the functional $J$ corresponding to the Navier--Stokes equation
and the energy input rate $I$. For the function space $X$ we set $X=L^2(\Omega)$
assuming that the state $\vc u$ belongs to the space of square integrable vector fields.
By definition, we have $J(\vc u)=dI(\vc u;\mathcal N(\vc u))$ which implies
\begin{align}
J(\vc u) & =\frac{1}{|\Omega|}\int_{\Omega}\left(-\vc u\cdot \bnabla\vc u -\bnabla p +\nu \Delta\vc u +\vc f\right)\cdot \vc f\id \vc x\nonumber\\
& = \frac{1}{|\Omega|}\int_{\Omega}\left[\vc u\cdot\left(\vc u\cdot\bnabla\vc f\right)+\nu\vc u \cdot \Delta \vc f\right]\id\vc x + \frac{1}{|\Omega|}\|\vc f\|^2_2,
\end{align}
where we used integration by parts. The term involving the pressure $p$ vanishes since the forcing is divergence free, 
$\bnabla\cdot \vc f=0$. Since $\|\vc f\|_2$ is constant, we can safely omit the second term and let
$$J(\vc u)=\frac{1}{|\Omega|}\int_{\Omega}\left[\vc u\cdot\left(\vc u\cdot\bnabla\vc f\right)+\nu\vc u \cdot \Delta \vc f\right]\id\vc x.$$

Next we compute the G\^ateaux differential of $J$. By definition, we have
\begin{align}
\id J(\vc u;\vc v) & = \frac{1}{|\Omega|}\int_{\Omega}\left[\vc v\cdot\left(\vc u\cdot\bnabla\vc f\right)+
\vc u\cdot\left(\vc v\cdot\bnabla\vc f\right)+\nu\,\vc v \cdot \Delta \vc f\right]\id\vc x\nonumber\\
& = \frac{1}{|\Omega|}\int_{\Omega}\left[\left( \bnabla\vc f + \bnabla\vc f^\top \right)\vc u +\nu \Delta\vc f\right]\cdot \vc v\,\id \vc x\nonumber.
\end{align}
On the other hand, by Riesz representation theorem, we have $\id J(\vc u;\vc v)=\langle J'(\vc u),\vc v\rangle_{L^2}$ which implies
\begin{equation}
J'(\vc u)=\frac{1}{|\Omega|}\left[\left( \bnabla\vc f + \bnabla\vc f^\top \right)\vc u +\nu \Delta\vc f\right].
\end{equation}

Similarly, the G\^ateaux differential of the constraint $C(\vc u)=\langle A\vc u,A\vc u\rangle_{L^2}/(2|\Omega|)$ is given by
\begin{equation}
\id C(\vc u;\vc v) = \frac{1}{|\Omega|}\langle A\vc u,A\vc v\rangle_{L^2} 
= \frac{1}{|\Omega|}\langle A^\dagger A\vc u,\vc v\rangle_{L^2} 
= \langle C'(\vc u),\vc v\rangle_{L^2},
\end{equation}
implying $C'(\vc u)=A^\dagger A\vc u/|\Omega|$. Finally, the adjoint of the divergence operator, $\mathcal K=\bnabla\cdot$, 
with respect to the $L^2$ inner product is the gradient operator, $\mathcal K^\dagger =-\bnabla$.
Substituting the above in the Euler--Lagrange equation~\eqref{eq:EL} and \eqref{eq:EL-const}, we obtain
\begin{subequations}
\begin{equation}
\left( \bnabla\vc f + \bnabla\vc f^\top \right)\vc u +\nu \Delta\vc f -\bnabla \alpha +\beta A^\dagger A\vc u=\vc 0,
\label{eq:uAlphaBetaPDE}
\end{equation}
\begin{equation}
\bnabla \cdot\vc u =0,
\label{eq:divFree}
\end{equation}
\begin{equation}
\frac{1}{|\Omega|}\int_{\Omega}\frac{|A \vc u|^2}{2}\id\vc x=c_0.
\label{eq:fixedE}
\end{equation}
\label{eq:opt_pde}%
\end{subequations}

A few remarks about equations~\eqref{eq:opt_pde} are in order:
(i) The PDE~\eqref{eq:uAlphaBetaPDE} 
is inhomogeneous due to the term $\nu \Delta\vc f=-\nu k_f^2\sin(k_f y)\vc e_1$. 
(ii)
The equations are nonlinear in the constraint~\eqref{eq:fixedE}. 
(iii) With the Kolmogorov forcing $\vc f = \sin(k_fy)\vc e_1$, 
the translations $\vc u(x,y)\mapsto \vc u(x+\ell ,y)$, with $\ell\in\mathbb R$, are a symmetry 
transformation of equations~\eqref{eq:opt_pde}. That is, if $\vc u(x,y)$ solves~\eqref{eq:opt_pde}, 
so does $\tilde{\vc u}(x,y)=\vc u(x+\ell,y)$ for all $\ell\in\mathbb R$.

\section{Newton iterations}\label{app:newton}
In this section, we outline the Newton iterations for solving the system~\eqref{eq:opt_pde}.
Define
\begin{equation}
\mathcal F(\vc u,\alpha,\beta)=
\begin{pmatrix}
\left( \bnabla \vc f + \bnabla \vc f^\top \right) \vc u +\nu \Delta \vc f -\bnabla \alpha +\beta A^\dagger A \vc u \\
\bnabla\cdot \vc u\\
\int_{\Omega}|A(\vc u)|^2\id \vc x-2|\Omega|c_0
\end{pmatrix}.
\end{equation}
The zeros of $\mathcal F$ coincide with the solutions of~\eqref{eq:opt_pde}. We find these zeros
numerically using damped Newton iterations
\begin{equation}
\vc u_{n+1}=\vc u_n+\epsilon\tilde{\vc u},\quad \alpha_{n+1}=\alpha_n+\epsilon\tilde\alpha,\quad \beta_{n+1}=\beta_n+\epsilon\tilde\beta.
\end{equation}
At each iteration, the Newton direction $(\tilde{\vc u},\tilde{\alpha},\tilde{\beta})$ is obtained as the solution of the linear equation
\begin{equation}
\mathcal L(\vc u_n,\alpha_n,\beta_n; \tilde{\vc u},\tilde{\alpha},\tilde{\beta})=-\mathcal F(\vc u_n,\alpha_n,\beta_n),
\label{eq:newton_dir}
\end{equation}
where $\mathcal L(\vc u,\alpha,\beta;\cdot,\cdot,\cdot)$ is the Gateaux differential of $\mathcal F$ 
at $(\vc u,\alpha,\beta)$ and is given explicitly as
\begin{equation}
\mathcal L(\vc u,\alpha,\beta; \tilde{\vc u},\tilde{\alpha},\tilde{\beta})=
\begin{pmatrix}
\left( \bnabla \vc f + \bnabla \vc f^\top \right) \tilde{\vc u} -\bnabla \tilde{\alpha} +\tilde{\beta} A^\dagger A \vc u+\beta A^\dagger A\tilde{\vc u} \\
\bnabla\cdot \tilde{\vc u}\\
2\int_{\Omega}A(\vc u)\cdot A(\tilde{\vc u})\id \vc x
\end{pmatrix}.
\end{equation}
The solution of the linear PDE~\eqref{eq:newton_dir} is approximated by 
the generalized minimal residual (GMRES) algorithm~\cite{gmres}.
At each iteration, the step size $\epsilon\in(0,1]$ is adjusted to achieve maximal 
decrease in the error $\|\mathcal F(\vc u_{n+1},\alpha_{n+1},\beta_{n+1})\|_{L^2}$ \cite{boyd04}.
The standard Newton iterations correspond to $\epsilon=1$.

\section{Sensitivity to parameters}
Recall that the constraint $\int_{\Omega}|\bnabla \vc u|^2\id\vc x/(2|\Omega|)=c_0$ enforces a constant energy dissipation rate. This constraint is motivated by the fact that, away from
extreme bursts, the energy dissipation rate $D$ exhibits small oscillations around its mean value. Nonetheless,
$D$ is not exactly constant, prompting the question whether the optimal solution is 
robust with respect to small perturbations to the constant $c_0$.
\begin{figure}
\centering
\includegraphics[width=.6\textwidth]{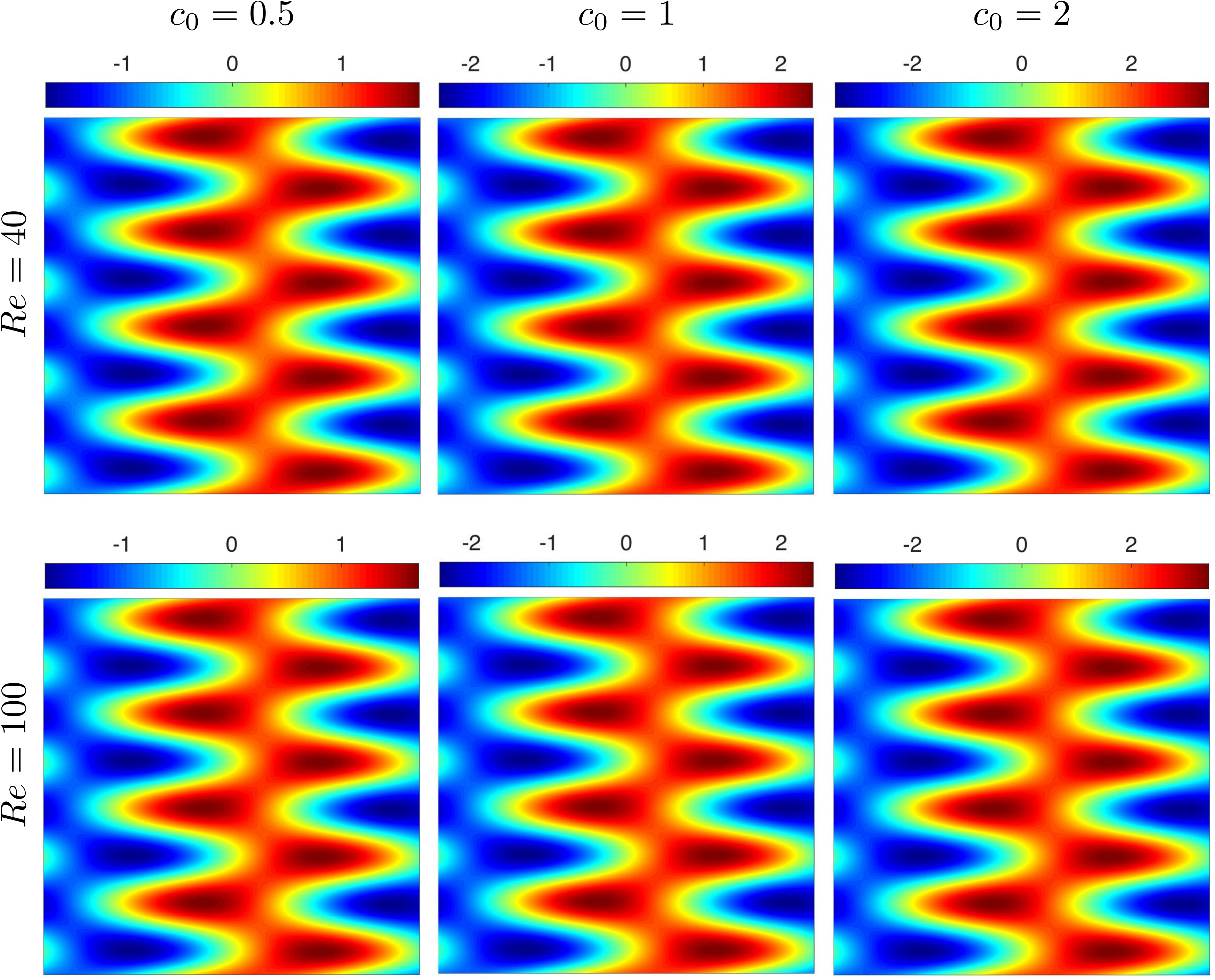}
\caption{The global optimal solutions with $c_0=0.5$, $1$ and $2$ at $Re=40$ and $Re=100$.}
\label{fig:c_0}
\end{figure}

To examine this robustness, we have computed the optimal solution for a wide range of parameters $c_0$.
We find that the optimal solution is in fact robust 
even with respect to relatively large variations in the parameter $c_0$.
Figure~\ref{fig:c_0}, for instance, shows the optimal solution for three different values of $c_0$ at $Re=40$ and $100$ 
(the results are similar for $Re=60$ and $80$). 

The insensitivity of the optimal solution with respect to the constant $c_0$ also implies that
the equality constraint $\int_{\Omega}|\bnabla \vc u|^2\id\vc x/(2|\Omega|)=c_0$
can be replaced with an inequality constraint of the form 
$c_1\leq \int_{\Omega}|\bnabla \vc u|^2\id\vc x/(2|\Omega|)\leq c_2$. For a wide range of values for $c_2>c_0>c_1>0$,
the optimal solutions corresponding to the two constraints will be similar.

\section{Computing the probability of extreme events}\label{app:Pee}
We approximate the conditional PDFs using the following steps.
For any two observables $\lambda$ and $\gamma$, we assume that their joint probability 
density function $p_{\gamma,\lambda}$ exists such that
\begin{equation}
\mathcal P(\gamma_1\leq \gamma\leq \gamma_2,\lambda_1\leq \lambda\leq \lambda_2)=
\int_{\gamma_1}^{\gamma_2}\int_{\lambda_1}^{\lambda_2}p_{\gamma,\lambda}(\gamma',\lambda')\id\lambda'\id \gamma'.
\end{equation}
Similarly, we also assume that the observable $\lambda$ has a probability density $p_\lambda$.
Once the PDF $p_\lambda$ and the joint PDF $p_{\gamma,\lambda}$ are approximated using direct numerical simulations,
the conditional PDF $p_{\gamma|\lambda}$ can be evaluated by the Bayesian formula, 
$$p_{\gamma|\lambda}=\frac{p_{\gamma,\lambda}}{p_{\lambda}}.$$

Computation of the extreme event probability $P_{ee}$
from the conditional probability is straightforward.
Let $\gamma_e$ denote the threshold such that $\gamma>\gamma_e$ denotes 
an extreme event. Then by definition, we have
\begin{equation}
P_{ee}\left(\lambda_0\right) = \mathcal P\left( \gamma>\gamma_e \big|\, \lambda=\lambda_0\right)
=\int_{\gamma_e}^{\infty}p_{\gamma|\lambda}(\gamma'|\lambda_0)\id \gamma',
\end{equation}
where $\gamma'$ is a dummy integration variable. 
In the present paper, the variable $\lambda$ is the indicator $|a(1,0)|$
and the variable $\gamma$ is the future maximum of the energy dissipation rate, 
$\gamma(t)=\Dmax(t)=\max_{\tau\in[t+t_i,t+t_f]} D(\vc u(\tau))$.
\edit{At each Reynolds number, the joint probability $p_{\gamma,\lambda}$ is approximated from the $100,000$
computed data points on a $20\times 30$ grid over the $(\gamma,\lambda)$ plane.}

\section{Supporting computational results}\label{app:results_highRe}
In this section, we present the numerical results for Reynolds numbers $Re=40$, $60$, $80$ and $100$.
The relevant parameters and variables are summarized in Table~\ref{tab:params}. At each Reynolds
number, the statistics are computed from long trajectory data of length $10,000$ time units. 
The states (i.e. the velocity fields $\vc u$) are saved along these 
trajectories at every $0.1$ time units, amounting to a combined $100,000$ distinct states at 
each Reynolds number. Before recording any data, we evolved random initial conditions 
for $500$ time units to ensure the decay of transients. 
%

\rowcolors{1}{gray!15}{white}
\begin{table}[h!]
\centering
\caption{Simulation parameters including 
the Reynolds number $Re$, 
the resolution $N\times N$,
the mean $\mathbb E[D]$ and the standard deviation
$\sqrt{\mathbb E[D^2]-\mathbb E[D]^2}$ of the energy dissipation rate $D$.
The eddy turn-over time $t_e$ and the prediction time $t_i$ are reported in terms of 
non-dimensional time units. The percentage of hits, correct rejections and the false positives and negatives
of the extreme event predictions are also reported.
The rate of successful predictions (RSP) and the rate of successful rejections (RSR) are 
computed from formula~\eqref{eq:rsp} and~\eqref{eq:rsr}.
}
\begin{tabular}{c||ccccccc}
$Re$ & 40 & 60 & 80 & 100 \\ \hline
$N$ & 128 & 256 & 256 & 256 \\ 
$\mathbb E[D]$ & 0.1168 & 0.1159 & 0.1010 & 0.0903 \\ 
$\sqrt{\mathbb E[D^2]-\mathbb E[D]^2}$ &  0.0384 & 0.0465  & 0.0369  & 0.0295 \\ 
$t_e$ & 0.46 & 0.38  & 0.35 & 0.33  \\ \hline
$t_i$  & 1.0 & 1.0 &  1.0 &  1.0 \\ 
$t_f$  & 2.0 & 2.0 &  2.0 &  1.5 \\ 
Hits & $5.60\%$ & $17.7\%$ & $15.3\%$ & $11.3\%$ \\
Correct Rejection & $93.3\%$ & $77.8\%$ & $78.5\%$ & $81.7\%$ \\
False Negatives & $0.26\%$  & $2.3\%$ &  $3.5\%$ & $4.3\%$ \\ 
False Positives &  $0.85\%$  &  $2.1\%$ &  $2.6\%$ & $2.6\%$ \\
RSP & 95.6\% & 88.4\% & 81.2\% & 72.3\% \\
RSR & 99.1\% & 97.4\% & 96.8\% & 96.9\%
\end{tabular}
\label{tab:params}
\end{table}
\begin{figure}
\centering
\includegraphics[width=.9\textwidth]{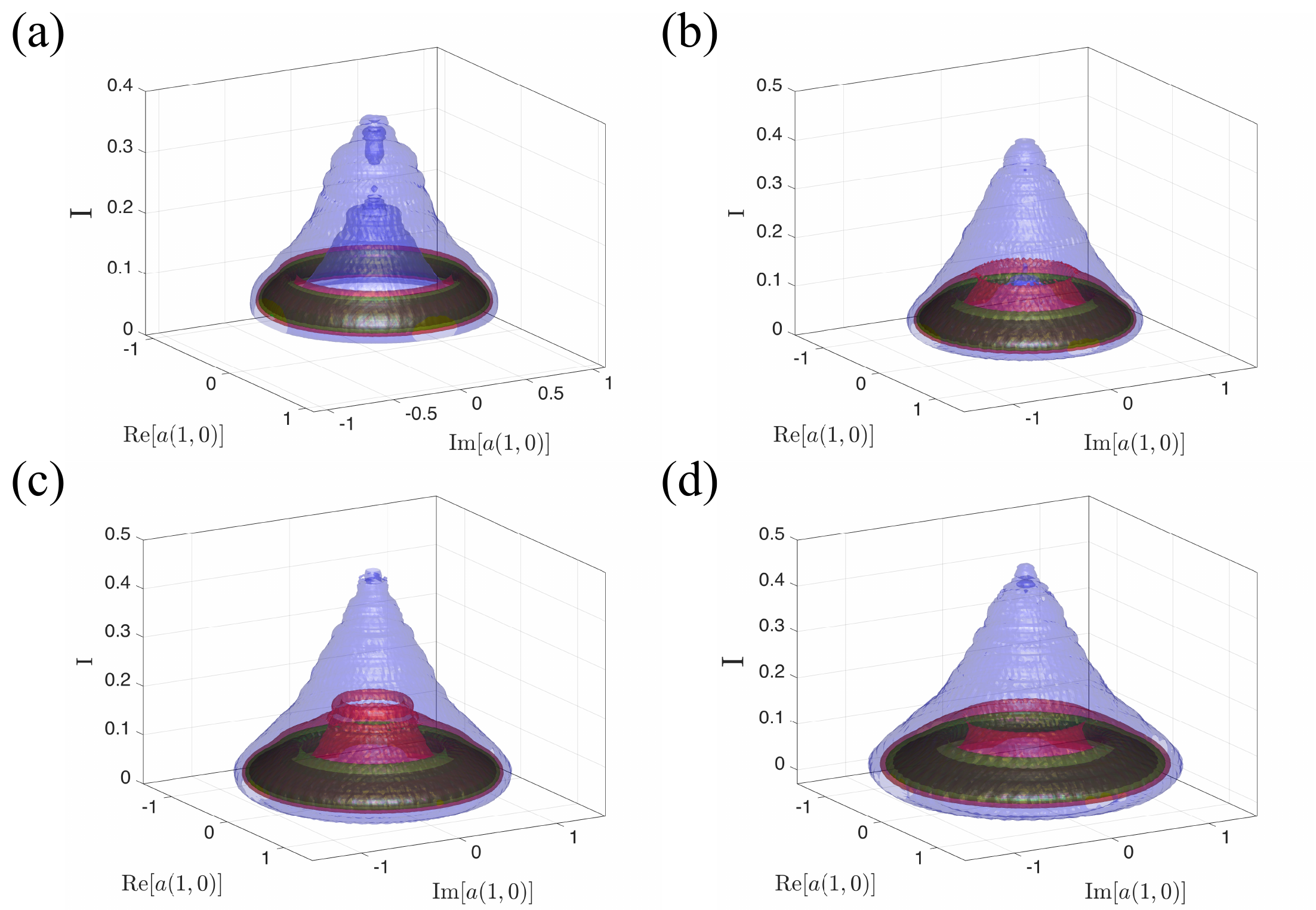}
\caption{The joint PDF of the energy input rate $I$, $\re[a(1,0)]$ and $\im[a(1,0)]$ at $Re=40$ (a), $Re=60$ (b) $Re=80$ (c)
$Re=100$ (d). The PDFs show that small values of $|a(1,0)|$ correlate strongly with the large values of the energy input rate $I$.
}
\label{fig:jointPDF}
\end{figure}

The Navier--Stokes equations are solved numerically with a standard pseudo-spectral code with $N\times N$
Fourier modes and 2/3 dealiasing~\cite{Fox1973} and a forth-order Runge--Kutta scheme for the temporal evolution.
For $Re=60$, $80$ and $100$, we use $256\times 256$ Fourier modes to fully resolve the velocity fields.
At $Re=40$, however, this resolution is unnecessarily high and hence we use $128\times 128$ modes.

Figure~\ref{fig:jointPDF} shows the joint PDFs of the mode $a(1,0)$ versus the energy input $I$.
At all Reynolds numbers the joint PDFs have a cone shape reflecting the fact that
small values of $|a(1,0)|$ correspond to large values of the energy input rate.
\begin{figure}[t]
\centering
\includegraphics[width=.9\textwidth]{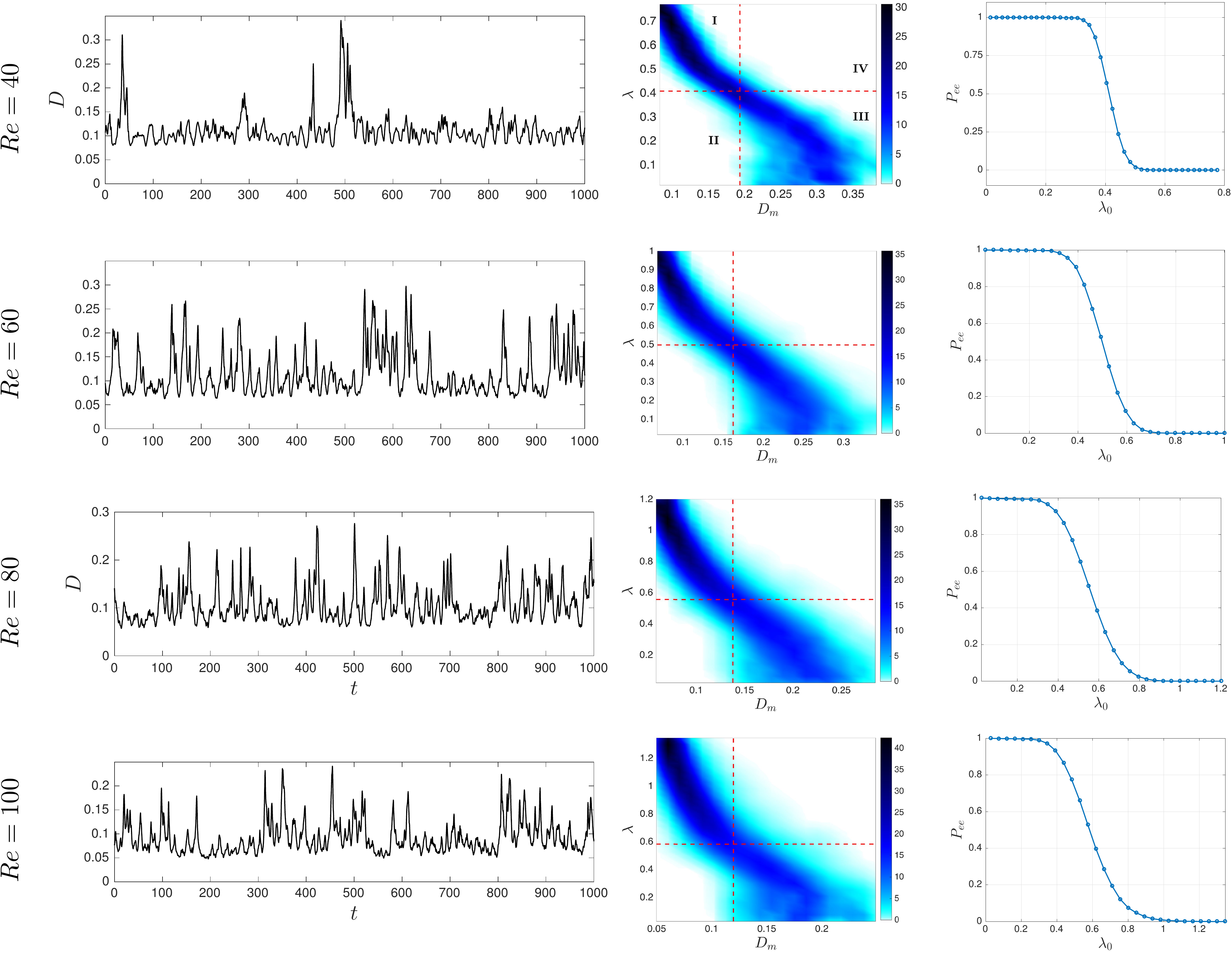}
\caption{The prediction of intermittent bursts of the energy dissipation rate $D$ at
Reynolds numbers $Re=40, 60, 80$ and $100$. 
Left column: Time series of the energy dissipation along a typical trajectory. 
Middle column: The conditional density $p(D_m|\lambda)$. 
Right column: Probability of extreme events $P_{ee}$.}
\label{fig:EngDiss}
\end{figure}

As in $Re=40$, we use the evolution of $|a(1,0)|$ to predict an upcoming burst of the energy dissipation $D$.
Figure~\ref{fig:EngDiss} shows the computational results at higher Reynolds numbers. 
For $Re=60$, $80$ and $100$, we set the threshold $D_e$ for the extreme energy dissipation 
to be the mean plus one standard deviation of the energy dissipation.
The measured mean $\mathbb E[D]$ 
and standard deviation $\sqrt{\mathbb E[D^2]-\mathbb E[D]^2}$ are reported in Table~\ref{tab:params}.
The corresponding extreme dissipation thresholds $D_e$ are marked by vertical red dashed lines
in the middle panel of figure~\ref{fig:EngDiss}. The horizontal dashed line marks the critical $\lambda_0$ 
for which $P_{ee}=0.5$, that is $50\%$ probability of an upcoming extreme event.

We recall from the main body of the paper that the four quadrants in the conditional PDFs (middle column of figure~\ref{fig:EngDiss})
correspond to:
\begin{enumerate}[(I)]
\item Correct rejection ($P_{ee}<0.5$ and $D_m(t)<D_e$): Correct prediction of no upcoming extremes.
\item False positives ($P_{ee}>0.5$ but $D_e(t)<D_e$): The indicator predicts an upcoming extreme event 
but no extreme event actually takes place.
\item Hit ($P_{ee}>0.5$ and $D_m(t)>D_e$): Correct prediction of an upcoming extreme.
\item False negatives ($P_{ee}<0.5$ but $D_m(t)>D_e$): An extreme event takes place but
the indicator fails to predict it.
\end{enumerate}

Table~\ref{tab:params} also shows the results of the extreme event prediction.
In order to quantify the success of these predictions, we define 
\begin{equation}
\mbox{Rate of Successful Predictions (RSP)}=\frac{\mbox{Hits}}{\mbox{Hits+False Negatives}},
\label{eq:rsp}
\end{equation}
which measures the ratio of the number of extreme events that were successfully predicted to the total number of 
extreme events. Similarly, the quantity,
\begin{equation}
\mbox{Rate of Successful Rejections (RSR)}=\frac{\mbox{Correct Rejections}}{\mbox{Correct Rejections+False Positives}},
\label{eq:rsr}
\end{equation}
measures the ratio of the number of non-extreme events that were correctly rejected to the total number of non-extreme events. 


\end{document}